\documentclass[letterpaper,twocolumn,10pt]{article}
\usepackage{zhanggroup}

\usepackage{tcolorbox}
\usepackage{xspace}
\usepackage{caption}
\usepackage{subcaption}
\usepackage{float}
\tcbuselibrary{breakable}

\newcounter{findingctr}
\newcommand{\finding}[1]{%
\refstepcounter{findingctr}%
\begin{tcolorbox}[
  colback=gray!10,
  colframe=black!60,
  boxrule=0.6pt,
  arc=3pt,
  left=6pt,
  right=6pt,
  top=6pt,
  bottom=6pt,
  width=\linewidth,
  breakable
]
\raggedright\tolerance=1000\emergencystretch=1em
\textbf{Finding \thefindingctr:~} #1
\end{tcolorbox}
}
\newenvironment{promptbox}{%
  \begin{tcolorbox}[
    colback=gray!6,
    colframe=black!70,
    boxrule=0.6pt,
    arc=3pt,
    left=8pt, right=8pt, top=6pt, bottom=6pt,
    width=\linewidth,
  ]
\ttfamily\small\raggedright
}{%
  \end{tcolorbox}
}
\AtBeginDocument{%
}

\newcommand{\mypara}[1]{\noindent{\bf {#1}.} \xspace}

\begin{document}
%-------------------------------------------------------------------------------

\date{}

\title{\bf From Celebrities to Anyone: Characterizing AI Nudification Content, Technology, and Community Dynamics on 4chan}

\author{
Chi Cui\textsuperscript{1}\ \ \
Yixin Wu\textsuperscript{1}\ \ \
Yang Zhang\textsuperscript{1}
\\
\\
\textsuperscript{1}\textit{CISPA Helmholtz Center for Information Security}
}

\maketitle

%-------------------------------------------------------------------------------
\begin{abstract}
AI nudification uses generative models to create synthetic non-consensual sexually explicit imagery (SNEACI) of real individuals.
Prior work has examined dedicated nudification platforms and model repositories, finding that most targets are female celebrities.
However, the anonymous content community, where SNEACI is actively requested, generated, and exchanged, remains unexplored.
In this work, we present a large-scale study of AI nudification in the wild, identifying 24,105 SNEACI items.
We find a significant shift in target demographics: non-celebrity individuals now account for 55.8\% of targets, compared to only 4.7\% in prior studies, indicating that AI nudification has expanded from targeting public figures to increasingly harming individuals within users' own social circles.
Meanwhile, open-source models dominate production, with Stable Diffusion family generating 42.7\% of images and Wan generating 66.5\% of videos, all driven by thousands of shared fine-tuned models and accessible tutorials.
Yet the ecosystem runs on a small cohort of active producers, with the most prolific producing 780 items, drives community engagement, shapes target demographics, and disseminates technical knowledge that lowers barriers for new producers.
Our work provides an empirical understanding of how AI nudification operates in the wild, revealing the mechanisms that sustain this ecosystem and highlighting the urgent need for interventions in platform governance, technical safeguards, and affected individual protection.
\end{abstract}
%-------------------------------------------------------------------------------

%-------------------------------------------------------------------------------
\section{Introduction}
\label{section:intro}
%-------------------------------------------------------------------------------

Synthetic non-consensual explicit AI-created imagery (SNEACI)~\cite{GOBCBTRK25} has emerged as one of the most pressing harms associated with generative AI.
Often referred to as ``AI nudification,'' this practice uses text-to-image, image-to-image, and video generation models to produce sexually explicit depictions of real individuals without their consent.
Notable incidents, such as the widely circulated AI-generated pornographic images of Taylor Swift in January 2024~\cite{DEEPFAKE_TAYLOR_SWIFT}, have drawn public attention, but the problem extends far beyond celebrities.
A more recent report indicates that an estimated 3 million sexualized images, including 23,000 of children, are generated using Grok~\cite{grok} over an 11-day period~\cite{GROK_SEXUALIZED_IMAGES_REPORT}.

The growing scale of this harm underscores the need to understand the ecosystems that produce and proliferate SNEACI.
Early works examine AI nudification communities including Reddit\footnote{Reddit's r/deepfakes was banned in 2018, for violating policies.} and MrDeepFakes~\cite{MRDEEPFAKES},\footnote{MrDeepFakes was closed on May 4, 2025.} finding that the tools available at the time, such as face-swap frameworks like DeepFaceLab~\cite{LPGCZZ23}, require substantial technical expertise~\cite{TMDGDGG23, HLKD25}.
As a result, only a small number of skilled producers could create such content, and the primary targets are female celebrities.
However, with the rise of diffusion-based image models (e.g., Stable Diffusion family~\cite{RBLEO22, PELBDMPR23}) and video generation models (e.g., Wan~\cite{WAWMXCYZYZWZZWCZZYHMZLWCFZSFWGWSLWWZWSYSHXKLLLWZHLWLPZHSFJHWL25}), the threat has escalated further.
Studies of AI nudification models~\cite{HMR25} and services~\cite{GOBCBTRK25} show how accessible the technology has become: users with no technical expertise can now generate SNEACI from even a single photograph of the target.
As the technical barrier declines, recent case studies~\cite{LD26,MQR26} suggest that AI nudification is increasingly targeting non-celebrities within users' own social circles, yet the ecosystems that enable this shift remain poorly understood.

In this work, we present the first large-scale measurement study of 4chan's \textit{Adult Requests} board, characterizing the request-fulfill SNEACI ecosystem through three lenses: the content it produces, the supply chain that enables production, and the ecosystem dynamics that sustain it.
We focus on 4chan for several reasons.
First, unlike MrDeepFakes,  which restricts content to celebrity targets and charges for customized services, 4chan represents a non-commercial community with no restrictions on the target population, directly complementing Han et al~\cite{HLKD25}.
Second, as a forum-based community, 4chan contains abundant observable request-fulfill behavior, making interaction between requester and provider available for analysis.
Third, its scale and accessibility make it amenable to systematic measurement.

\mypara{Research Questions}
We aim to provide a full characterization of the SNEACI ecosystem on 4chan by addressing the following research questions:
\begin{itemize}
    \item \textbf{RQ1: What does SNEACI content look like on 4chan?}
    We characterize SNEACI content by media type and target distribution, distinguishing between celebrity and non-celebrity targets.
    \item \textbf{RQ2: What supply chain supports SNEACI production?}
    We trace the provenance of SNEACI to specific generative models and examine the resources circulated within the community, including fine-tuned models, tutorials and external platforms.
    \item \textbf{RQ3: What are the ecosystem dynamics of the community?}
    We analyze the social structure of the request-fulfill ecosystem on 4chan, characterizing interaction patterns between requesters and providers, and identifying the core actors who sustain community activity.
\end{itemize}
To address these questions, we track the community's activity over a 41-day period, collecting and analyzing 3,661 threads including 80,366 posts containing images and videos.
To identify SNEACI from collected media, we develop a multi-stage detection framework (\autoref{figure:pipeline}).
We begin by classifying all collected images through an NSFW classifier to retain only sexually explicit content.
NSFW images are then passed through an AIGC detector to identify fully synthetic material, followed by a custom undress detector to capture AI-nudified images where only specific body regions are modified.
Manual inspection of a random sample of 200 videos confirm that over 98\% are AI-generated sexual content.
Given the negligible proportion of non-SNEACI videos and the impracticality of manually verifying all 8,203 videos, we treat all successfully processed videos as SNEACI.
Finally, we classify all identified SNEACI by target population, distinguishing between celebrity and non-celebrity individuals.

\mypara{Key Insights}
We summarize our key insights as follows:

\noindent \textit{RQ1: SNEACI content.}
Video SNEACI is emerging as an increasingly significant modality, driven by the maturation of short video generation and a substantial reduction in technical barriers.
AI nudification has shifted from a celebrity-centric practice to one increasingly targeting ordinary individuals: non-celebrities account for 55.78\% of all SNEACI targets, rising to 60.26\% in video content.
This shift has been systematically overlooked in prior work due to three visibility biases: the dispersion of non-celebrity targets, the limited dissemination of non-celebrity content, and the lack of public attention.
The shift itself is driven by two forces: the reduction of technical barriers that once made celebrities the most accessible targets, and a request-fulfill dynamic that directly surfaces end users' intent to target people within their own social circles.

\noindent \textit{RQ2: Supply chain.}
Open-source models lacking safety guardrails form the production backbone of the SNEACI supply chain: the Stable Diffusion family powers 42.4\% of images and Wan accounts for 66.5\% of videos.
The supply chain further sustains production through shared NSFW fine-tuned models, which overwhelmingly target female celebrities and remain difficult to permanently remove from the Internet, and a network of external platforms spanning capability provision, content distribution, and off-platform migration.

\noindent \textit{RQ3: Ecosystem dynamics.}
The request-fulfill ecosystem is predominantly oriented toward non-celebrity targets, with non-celebrity requests outnumbering celebrity requests nearly threefold and accounting for more than twice the volume of SNEACI content, underscoring the substantial and largely overlooked risk faced by ordinary individuals.
This ecosystem is sustained by a small core of active providers who function as ecosystem architects: they catalyze request volume, exploit requesters to amplify harm against targets, and cultivate new providers, ensuring the ecosystem's self-perpetuation.

\mypara{Implications}
Our findings carry implications across multiple fronts.
For platform governance, the multi-platform nature of the SNEACI supply chain suggests that content moderation on any single platform is insufficient; effective intervention requires coordinated action across resource-sharing, content-delivery, and private community platforms.
For the open-source AI community, the dominance of models lacking safety guardrails underscores the urgent need for more robust safeguards against misuse.
Most critically, the prevalence of non-celebrity targets reveals that AI nudification is no longer a problem confined to public figures.
Ordinary individuals also face substantial and largely unrecognized risks, demanding greater attention from both researchers and policymakers.

%-------------------------------------------------------------------------------
\section{Background and Related Work}
\label{section:background}
%-------------------------------------------------------------------------------

%-------------------------------------------------------------------------------
\subsection{Image and Video Generation Techniques}
\label{section:bg_generation_tech}
%-------------------------------------------------------------------------------

The capabilities of generative AI for image and video synthesis have advanced rapidly in recent years.
Early image generation models had relatively limited capabilities.
For example, Generative Adversarial Network (GAN) could generate synthetic images, but often struggled to produce high-quality, controllable, and highly realistic outputs.
The introduction of diffusion-based text-to-image models fundamentally transformed this landscape.
Models such as Stable Diffusion~\cite{RBLEO22}, SDXL~\cite{PELBDMPR23}, and Flux~\cite{flux} can generate photorealistic images from text prompts and support fine-tuning methods for further customization.
A representative example is Low-Rank Adaptation (LoRA)~\cite{HSWALWWC22}, which updates only a small set of additional parameters while keeping the base model frozen.
The resulting LoRA weights can then be combined with the base model at inference time, enabling users to generate images with specific attributes, such as a particular style or object.
Video generation has recently reached a similar level of accessibility.
Open-source models such as Wan~\cite{WAWMXCYZYZWZZWCZZYHMZLWCFZSFWGWSLWWZWSYSHXKLLLWZHLWLPZHSFJHWL25} and Hunyuan~\cite{KTZMDZXLWZWLYLWWLHYTWSBWXWWLLLWYDLCCPYHXZXTLLZWYWLJZ24} support image-to-video workflows that can animate a single reference photograph into a short video clip.
Moreover, public repositories such as Civitai and Hugging Face host thousands of publicly accessible resources (e.g., LoRAs) that further enhance these models for specific effects and applications.

Despite their impressive capabilities, these models exhibit significant safety gaps.
Previous studies~\cite{QSHBZZ23, PXZW25, LSHDQLSS25} show that these models can be prompted to generate unsafe content, including hateful memes, violent imagery, and sexually explicit material.
In this paper, we focus on one particularly harmful form of misuse: the generation of sexual imagery involving real individuals.

%-------------------------------------------------------------------------------
\subsection{AI Nudification}
\label{section:bg_nudification}
%-------------------------------------------------------------------------------

AI nudification refers specifically to the use of generative AI to produce non-consensual sexually explicit imagery of real individuals.
Timmerman et al.~\cite{TMDGDGG23} conducted an early study of AI nudification communities on both Reddit and MrDeepFakes~\cite{MRDEEPFAKES}, a prominent SNEACI marketplace at the time.
Later, Han et al.~\cite{HLKD25} further analyzed the MrDeepFakes ecosystem, examining its economics, target demographics, and user discussions.
However, the technology in that period are still dominated by face-swap tools such as DeepFaceLab~\cite{LPGCZZ23}, which require substantial technical expertise and are limited to skilled users.

The advent of diffusion models has dramatically expanded the scope and scale of AI nudification, and the increasing frequency of such incidents~\cite{DEEPFAKE_NEWS_UNWOMAN, DEEPFAKE_Guardian_Dan, DEEPFAKE_Guardian_Sally} has prompted a new wave of research.
Hawkins et al.~\cite{HMR25} measure the proliferation of deepfake text-to-image model variants on Hugging Face and Civitai, revealing the existence of explicit sexual models.
They find that Stable Diffusion family and Flux are the dominant base architectures, and that identity-specific LoRAs have become the primary mechanism for targeting individuals.
These LoRAs are trained on images of a specific individual, enabling users to combine them with a base model at inference time to generate images resembling that person.
Unlike earlier technologies, these models substantially lower the barrier to producing SNEACI, enabling the generation of a specific individual with as few as 20 photographs.
Gibson et al.~\cite{GOBCBTRK25} conduct the first systematic audit of commercial end-to-end nudification services, examining 20 websites that offer nudification as a paid service.
They find that these websites provide features ranging from undressing to rendering subjects in sexual positions.
Critically, these services require virtually no technical expertise from users, needing only an uploaded photograph and a click.
Together, recent studies illuminate the commercial platforms and model repositories that package AI nudification for end users, showing that the barrier to producing such content has become remarkably low while the underlying capabilities have grown substantially.
However, they do not directly observe the content being produced and exchanged in the wild.

In this work, we conduct a large-scale empirical study of AI nudification on 4chan.
Compared with MrDeepFakes, which is centered on creators actively uploading videos and explicitly restricts targets to celebrities, 4chan is mainly driven by anonymous user requests and imposes no explicit restrictions on the target population.
By investigating 4chan \textit{Adult Requests}, our work examines what SNEACI looks like in practice, how it is produced, and how user interactions shape its demand and supply.

%-------------------------------------------------------------------------------
\section{RQ1: Characterizing SNEACI Content}
%-------------------------------------------------------------------------------

Prior work has largely focused on SNEACI content generation techniques predating the diffusion model era, where technical barriers were substantially higher~\cite{TMDGDGG23, HLKD25}.
The rapid advancement of diffusion-based models has since dramatically lowered the required technical expertise, dependency on target-specific images~\cite{HMR25}, and API access costs~\cite{GOBCBTRK25}, while video generation is increasingly becoming feasible~\cite{WAWMXCYZYZWZZWCZZYHMZLWCFZSFWGWSLWWZWSYSHXKLLLWZHLWLPZHSFJHWL25}.
Therefore, prior findings~\cite{TMDGDGG23, HLKD25} may no longer hold, underscoring an urgent need to investigate the current landscape of SNEACI content in the wild.

\begin{figure}
  \centering
\includegraphics[width=\linewidth]{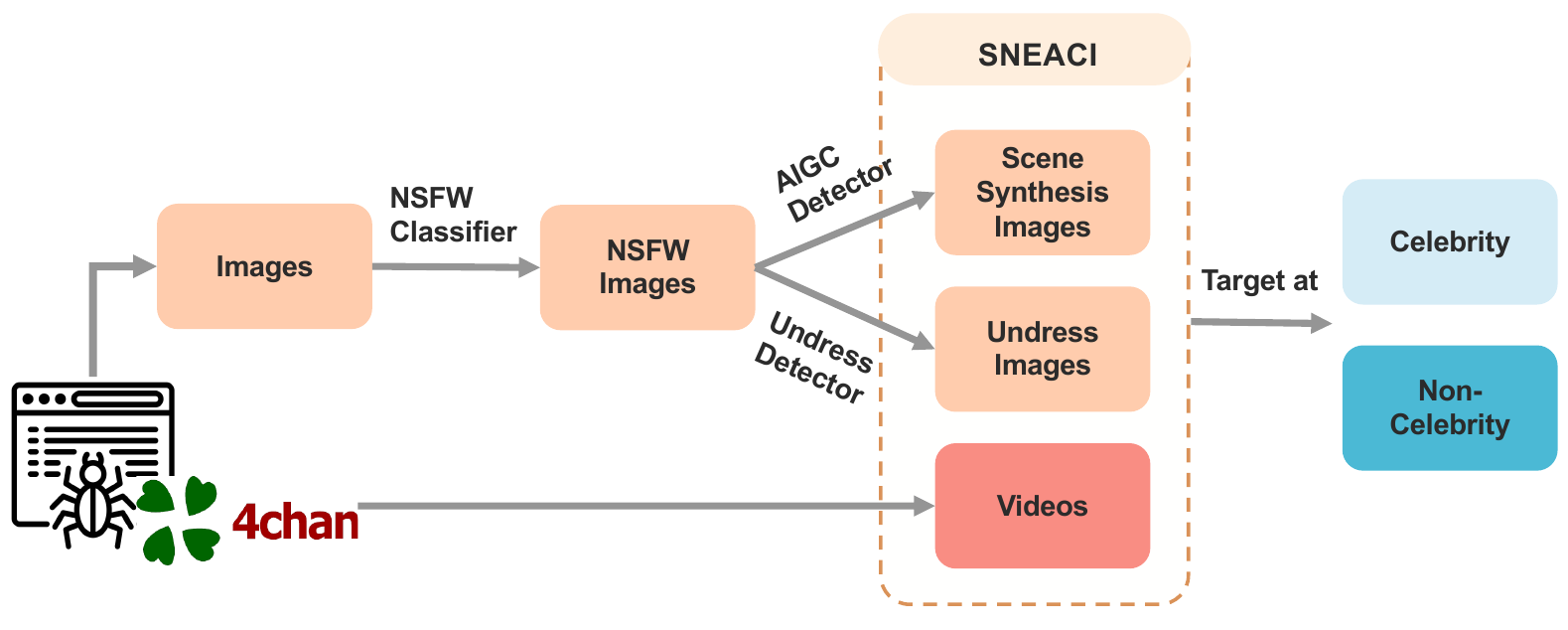}
  \caption{Multi-stage detection pipeline for identifying SNEACI from 4chan media files.}
\label{figure:pipeline}
\end{figure}

%-------------------------------------------------------------------------------
\subsection{Data Collection}
\label{section:rq1_data_collection}
%-------------------------------------------------------------------------------

Specifically, we focus on 4chan's \textit{Adult Requests} board, which serves as a primary venue for such content.
We develop a custom crawler to scrape archived threads from this board over the period from January 27 to March 08, 2026.
From the collected threads, we extract image and video files along with their associated metadata, including post time, thread ID, post author, and post content.
In total, we collect 3661 threads, including 80,366 posts with 49,874 media files.

%-------------------------------------------------------------------------------
\subsection{Data Processing}
\label{section:rq1_data_processing}
%-------------------------------------------------------------------------------

\mypara{Overview}
The collected dataset consists of two primary media types: images and videos.
The \textit{Adult Requests} board primarily serves as a venue for exchanging AI-generated nudification content.
Manual inspection of a random sample of 200 videos confirm that over 98\% are AI-generated sexual content.
Given the negligible proportion of non-SNEACI videos and the impracticality of manually verifying all 8,203 videos, we treat all successfully processed videos as SNEACI.
For images, which include a mix of source photos, memes, and AI-generated content, we apply a multi-stage pipeline, as illustrated in~\autoref{figure:pipeline}.
We first apply an \textit{Not-Safe-for-Work (NSFW) Detector} to retain only sexually explicit images, discarding Safe-for-Work (SFW) content (e.g., source images posted as requests and memes).
For the remaining NSFW images, we use an \textit{AI-Generated Content (AIGC) Detector} to determine whether they are AI-generated.
Images classified as AI-generated are treated as SNEACI, as they depict targets placed in entirely synthetic sexual scenes.
However, certain AI-nudified images may evade the AIGC Detector.
This occurs particularly in undressing scenarios, where only specific body regions are modified while the background and face remain authentic, causing the majority of the image to appear unaltered.
To address this, we apply a customized \textit{Undress Detector} to images classified as real by the AIGC detector.
Images identified as undressed are also categorized as SNEACI.
Having identified SNEACI through the above pipeline, we further categorize them based on their targets.
We apply a \textit{Celebrity Classifier} to categorize each piece of SNEACI based on whether it depicts a celebrity or a non-celebrity.

\mypara{Detectors and Classifiers}
In the above process, we apply four specialized detectors and classifiers, each targeting a distinct stage of the pipeline.
Below, we describe each component and how it is applied.
Implementation details of each detector are provided in~\autoref{appendix:implementation_details_of_detectors}.

\begin{itemize}
\item \textbf{NSFW Detector.} 
To figure out whether an image contains sexual content, we apply a binary NSFW detector to filter out safe-for-work images.
\texttt{Falconsai/nsfw\_image\_detectio
n}~\cite{nsfw_detector} is a binary classifier for NSFW content filtering, and has been widely used in prior studies to detect unsafe content~\cite{ZRZGB25, NKKPSJLKM25, CWGMGZ25}.
We therefore use it to efficiently conduct an initial screening of our large-scale dataset.

\item \textbf{AIGC Detector.}
To separate AI-generated sexual images from real sexual photographs, we apply an AIGC detector.
Prior studies have shown that the commercial AIGC detector Hive AI~\cite{HIVE_AI_CONTENT_DETECTION} achieves the best automated performance in distinguishing AI-generated images from real images~\cite{HPBSSZZ24}.
Therefore, we adopt Hive AI to determine whether an image is AI-generated.
For each image, the API returns a confidence score for \texttt{ai\_generated} vs.\ \texttt{real} classification.
We classify an image as AI-generated when the \texttt{ai\_generated} confidence exceeds 0.5.
In addition to the binary classification result, the API also provides a ranked list of potential source models with confidence scores.
We use this information to analyze the provenance of SNEACI in~\autoref{section:rq2_provenance}.

\item \textbf{Undress Detector.}
Because the AIGC detector is designed to identify fully AI-generated images, it is not effective at detecting undress image, where only partial regions (e.g., clothing areas) have been synthetically modified while the remainder of the photograph is authentic.
To address this gap, we fine-tune a dedicated binary classifier specifically for undress detection.
We use a \texttt{ViT-Base-Patch16-224}~\cite{vit_base_patch16_224} architecture, fine-tune it on a manually curated dataset of 206 images (real photographs and undress-edited counterparts sourced from the collected 4chan data).
On the validation set, the detector achieves an F1 score of 1.0 for the undress class.

\item \textbf{Celebrity Classifier.}
Before introducing the celebrity classifier, we first clarify our definition of celebrities.
In this paper, we use celebrity to refer broadly to public figures with substantial public visibility and social influence.
This includes traditional celebrities, such as actors, singers, politicians, and athletes, as well as online influencers.
Following Hawkins et al.~\cite{HMR25}, we treat influencers with more than 10,000 followers on a single platform (e.g., TikTok or Instagram), as celebrities.

Distinguishing celebrities from non-celebrities is challenging.
Because the scope of celebrity is broad and dynamic: some individuals may be well known only within specific communities and unfamiliar to others, and celebrity status change over time.
Existing celebrity classifiers are typically trained on predefined identity datasets, which limits the range of celebrities they can recognize~\cite{CElEBRITY_CLASSIFIER}.
We therefore use Gemini as the backbone of our classifier, as it provides stronger visual understanding capabilities than other large language models (LLMs)~\cite{LLM_IMAGE_RANKING} and it supports face recognition.
However, our initial investigation finds that using a single Gemini leads to a high false positive rate, as it may misidentify ordinary individuals as celebrities.
We therefore design a dual-agent system driven by \texttt{gemini-3-flash-preview}~\cite{gemini_3_flash}, where a second auditing agent verifies the prediction of the first agent before assigning a celebrity label.
This approach achieves an accuracy of 88\% on a manually labeled validation set of 100 images compared with the celebrity classifier trained on a fixed dataset~\cite{CElEBRITY_CLASSIFIER} that only shows a 68\% accuracy.
\end{itemize}

%-------------------------------------------------------------------------------
\subsection{Results}
%-------------------------------------------------------------------------------

Through this pipeline, we identify a total of 15,902 SNEACI images: 13,539 detected by the AIGC detector and an additional 2,363 identified by the undress detector.
Combined with the 8,203 videos, our final dataset contains 24,105 SNEACI files.
We now report key characteristics of this dataset across two dimensions: media type composition and target distribution.

%-------------------------------------------------------------------------------
\subsubsection{Types of SNEACI}
%-------------------------------------------------------------------------------

\begin{figure}
\centering
\includegraphics[width=0.8\linewidth]{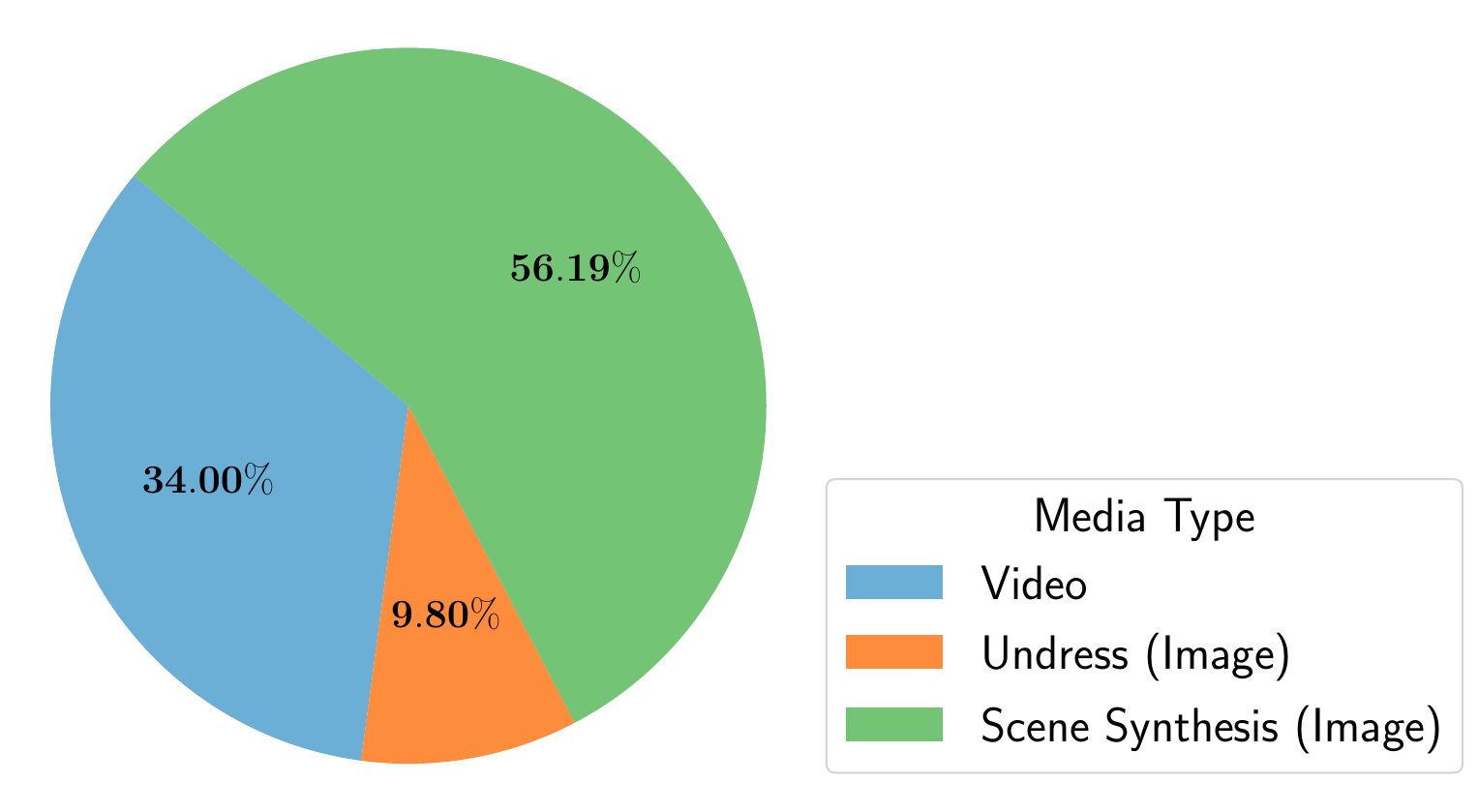}
\caption{Distribution of SNEACI by content type: video, scene synthesis (image), and undress (image).}
\label{figure:media_type_distribution}
\end{figure}

We classify the identified SNEACI into three categories based on format and generation technique.

\begin{itemize}
\item \textbf{Video}: AI-generated video clips depicting sexual content featuring the target individual.
These are typically generated from a benign input image and a text prompt, which guide the model to place the depicted person into a sexual scene and produce explicit actions.
\item \textbf{Scene Synthesis (Image)}: Fully AI-generated images where the target is placed into a synthetic sexual scene.
\item \textbf{Undress (Image)}: Partially modified images where AI selectively removes or replaces clothing of the target within an authentic photograph while leaving the rest of the image unchanged.
\end{itemize}

As shown in~\autoref{figure:media_type_distribution}, video accounts for 34.03\% of the total SNEACI in our dataset.
The remaining 65.97\% consists of images, with scene synthesis images and undress images.
This distribution fills an important gap in prior findings.
Prior studies of deepfakes focused exclusively on images~\cite{HMR25}, leaving video underexplored.
The substantial share of video in our dataset indicates that video generation capabilities have matured to the point where video has become a non-negligible medium for SNEACI.
Further, we find short videos of 5.06--8.06 seconds dominate the video content (details in~\autoref{appendix:durations_of_video_content}).

Scene synthesis images remain the single largest category.
This is likely because image generation is more accessible and less resource-intensive than video generation.
Taking the Wan2.7 series as an example, API-based video generation costs approximately 17 $\times$ more per clip than image generation~\cite{wan_pricing}.
The resource gap also exists for local deployment.
Image generation models such as SDXL require 8GB VRAM, whereas video generation models such as Wan2.1 require 24GB, which remains out of reach for many users, keeping images the preferred modality for the majority of producers.

\finding{While AI-generated images remain the dominant format, video SNEACI is emerging as an increasingly significant modality, driven by the maturation of short video generation and a substantial reduction in technical barriers.}

\begin{table}[!t]
\centering
\caption{Target distribution of SNEACI by media type.}
\label{table:target_distribution}
\scalebox{0.7}{
\begin{tabular}{l | c | c}
\toprule
& Celebrity & Non-Celebrity\\
\midrule
Images &  46.53\%\ & 53.47\%\\
Videos & 39.74\% & 60.26\%\\
\midrule
Total & 44.22\%  & 55.78\% \\
\bottomrule
\end{tabular}}
\end{table}

%-------------------------------------------------------------------------------
\subsubsection{Target Distribution}
\label{section:rq1_targets}
%-------------------------------------------------------------------------------

Using our celebrity classifier, we categorize each piece of SNEACI by whether it depicts a celebrity or non-celebrity.
As shown in~\autoref{table:target_distribution}, 55.78\% of SNEACI target non-celebrities, while 44.22\% target celebrities.
Additionally, videos are more likely to target non-celebrities than images, with 60.26\% of videos depicting non-celebrities compared to 53.47\% for images.
This indicates that non-celebrities, compared to celebrities, are more likely to be targeted by AI nudification, especially in videos.
This is particularly concerning, as video content is more realistic and therefore more harmful to affected individuals.
These findings challenge the prevailing assumption in prior work that AI nudification primarily targets celebrities.
For example, Han et al.~\cite{HLKD25} study a sexual deepfake marketplace whose products primarily target celebrities, and Hawkins et al.~\cite{HMR25} focus on celebrity-targeted deepfake models on public hosting platforms.
Yet our data show that non-celebrities account for the majority of SNEACI in the wild, suggesting that the threat is far more intimate and socially proximate than previously recognized.

\mypara{Why Non-Celebrities Are Overlooked as Targets}
We argue that this underrecognition is shaped by at least three visibility biases:

\begin{itemize}
\item \textbf{Dispersion of Targets.}
Celebrity targeting is highly concentrated, while non-celebrity targeting is widely dispersed.
In our dataset, we find 2,700 distinct celebrities, and the top 10 most targeted celebrities account for 15.71\% of all celebrity-focused content, and the top 50 account for 38.35\%.
This concentration makes individual celebrity cases highly salient.
By contrast, non-celebrity targeting is dispersed across a far larger number of individuals, each appearing in only a small number of items.
As a result, even if non-celebrities constitute the majority, no single case rises to public visibility.

\item \textbf{Limited Dissemination.}
Celebrity-targeted content is more likely to circulate on widely accessible platforms due to public recognition, whereas non-celebrity content is more covert and may go unnoticed even by the targets themselves.
Prior studies of sexual deepfake marketplaces~\cite{HLKD25} and model repositories~\cite{HMR25} show that celebrity-targeted products and models disseminate on public platforms, while content targeting non-celebrities are less visible.
We also observe this pattern in our dataset.
We find SNEACI targeting the same celebrity appearing in 38 different threads.
In comparison, content targeting non-celebrities rarely leaves the originating thread, making it invisible to studies that sample from widely accessible platforms.

\item \textbf{Lack of Public Attention.}
Celebrity incidents are more likely to receive media coverage and public attention, and celebrities often have greater resources and support to respond.
Non-celebrities usually lack similar channels and may also hesitate to come forward due to privacy concerns, leading to substantial underreporting.
\end{itemize}

\finding{Non-celebrities constitute the majority of SNEACI targets, especially in video content (60.26\%), yet remain systematically overlooked in prior work due to three visibility biases: the dispersion of non-celebrity targets, the limited dissemination of non-celebrity content, and the lack of public attention.
}

\mypara{AI Nudification Is Moving into Intimate Social Contexts}
Our results show that non-celebrities account for the majority of targets in AI nudification, suggesting that this practice is no longer primarily centered on public figures.
A necessary condition for this shift is the rapid advancement of AI image and video generation technologies, which has substantially lowered the technical barriers.
Previously, Hawkins et al.~\cite{HMR25} show that fine-tuning a personalized sexual image generator requires as few as 20 clear photographs of the target.
This made celebrities, with their abundant publicly available images, the most accessible targets.
However, our findings suggest that current AI nudification techniques can be carried out using only a single image.
For undressing, a fine-tuned Stable Diffusion model can directly take a single photograph as input and produce a nudified version.
For video generation, users can similarly provide a single photograph as a reference frame and use text prompts to guide the model in placing the target into a sexual scene and animating explicit actions, which explains why videos tend to target non-celebrities more than images.
Furthermore, several AI nudification applications offer nudification services at prices as low as \$0.06 per image~\cite{GOBCBTRK25}, making this form of abuse nearly costless and effectively barrier-free.
This is particularly concerning because, as AI nudification becomes easier to use, it is no longer limited to people with technical expertise and can instead be used by ordinary individuals to harm others in their social circles, making virtually anyone a potential target.

Our requester-side observations further reinforce this shift toward intimate and socially proximate targeting.
As we analyze in detail in~\autoref{sec:rq3}, the \textit{Adult Requests} board operates through a request-fulfill dynamic in which users who seek SNEACI (requesters) post requests and those with generation capabilities (providers) respond.
Prior studies have mainly examined marketplaces (e.g., MrDeepfake)~\cite{HLKD25,GOBCBTRK25} and model-sharing platforms (e.g., Hugging Face)~\cite{HMR25}, where the available content is largely determined by what providers choose to publish.
On such platforms, providers tend to publish celebrity-targeted content and models, which often receive greater attention and achieve wider dissemination.
In contrast, the request-fulfillment model on \textit{Adult Requests} board allows requesters to explicitly post their requests and wait for providers to respond.
This setting more directly reveals who end users actually intend to target, including acquaintances or friends~\cite{MQR26, LD26}, showing that AI nudification extends beyond public figures into more intimate and socially proximate contexts.

\finding{The shift in AI nudification targets is driven by two forces: the elimination of technical barriers that once made celebrities the most accessible targets, and a request-fulfill community that directly surfaces end users' intent to target people within their own social circles.}

%-------------------------------------------------------------------------------
\section{RQ2: Characterizing the Supply Chain}
%-------------------------------------------------------------------------------

While previous work has shown that a large number of AI nudification model repositories are available online~\cite{HMR25}, it remains poorly understood how these models are actually adopted to produce SNEACI and circulated in the wild.
In this section, we therefore first trace the provenance of SNEACI to identify which models are commonly used in practice.
We then analyze the resources shared within the community, including customized models, generation techniques, and various external platforms, to characterize the supply chain that enables SNEACI production.

%-------------------------------------------------------------------------------
\subsection{Provenance of SNEACI}
\label{section:rq2_provenance}
%-------------------------------------------------------------------------------

%-------------------------------------------------------------------------------
\subsubsection{Methodology}
%-------------------------------------------------------------------------------

To ensure methodological consistency with our AIGC detection pipeline, we use Hive AI to identify the generative model responsible for each piece of SNEACI.
Since undress images involve editing real photographs rather than generating images from scratch, they are often misclassified as real by AIGC detectors, making model attribution unreliable.
We therefore limit model attribution to scene synthesis images and videos.
For scene synthesis images, Hive AI flags the content as AI-generated and returns a ranked list of candidate source models with confidence scores; we assign each image to the top-ranked model.
For videos, we randomly sample five frames, run the Hive AI detector on each frame, and assign the video to the model selected by majority vote across the frame-level predictions.

%-------------------------------------------------------------------------------
\subsubsection{Results}
%-------------------------------------------------------------------------------

We report the distribution of source models for both images (only scene synthesis) and videos.
We further break down the results by target type (celebrity vs.\ non-celebrity) to examine whether different models are associated with different target groups.

\begin{figure*}
\centering
\begin{minipage}[c]{0.78\textwidth}
\begin{subfigure}[b]{0.25\linewidth}
\centering
\includegraphics[width=\linewidth]{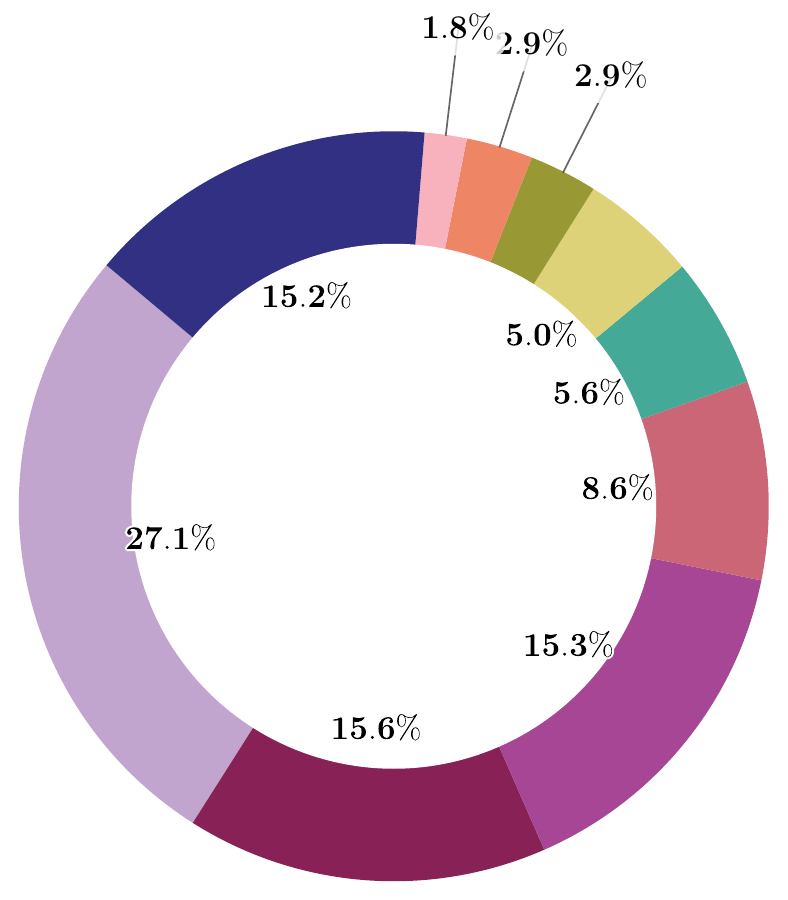}
\caption{Image (All Targets)}
\label{figure:all_image_models_donut}
\end{subfigure}
\hfill
\begin{subfigure}[b]{0.25\linewidth}
\centering
\includegraphics[width=\linewidth]{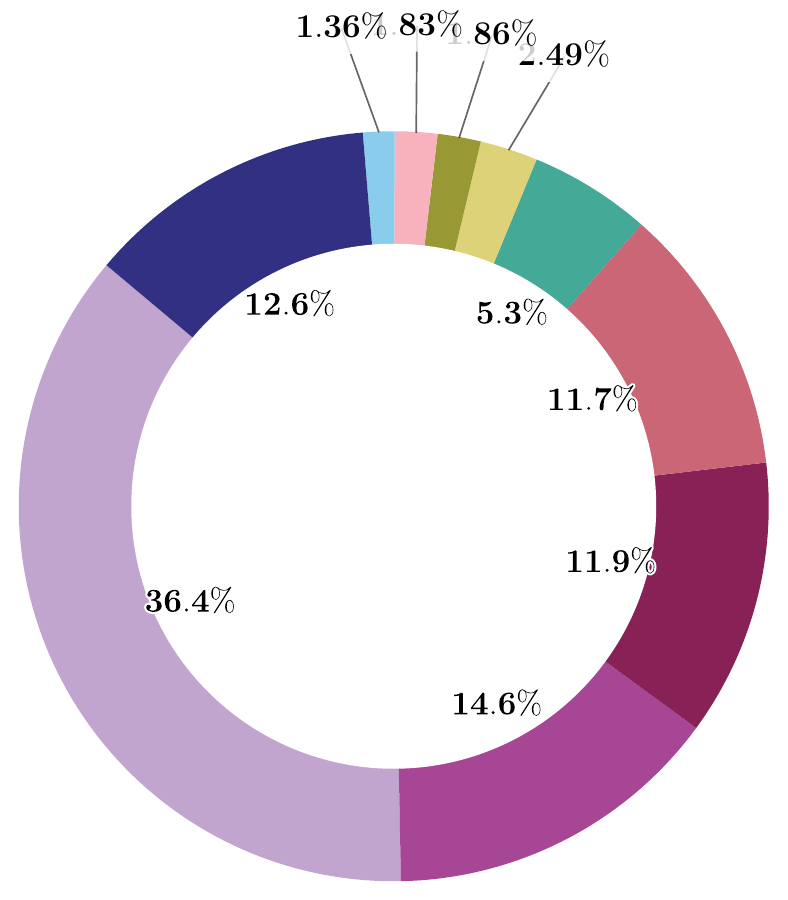}
\caption{Image (Celebrity)}
\label{figure:celebrity_image_models_donut}
\end{subfigure}
\hfill
\begin{subfigure}[b]{0.25\linewidth}
\centering
\includegraphics[width=\linewidth]{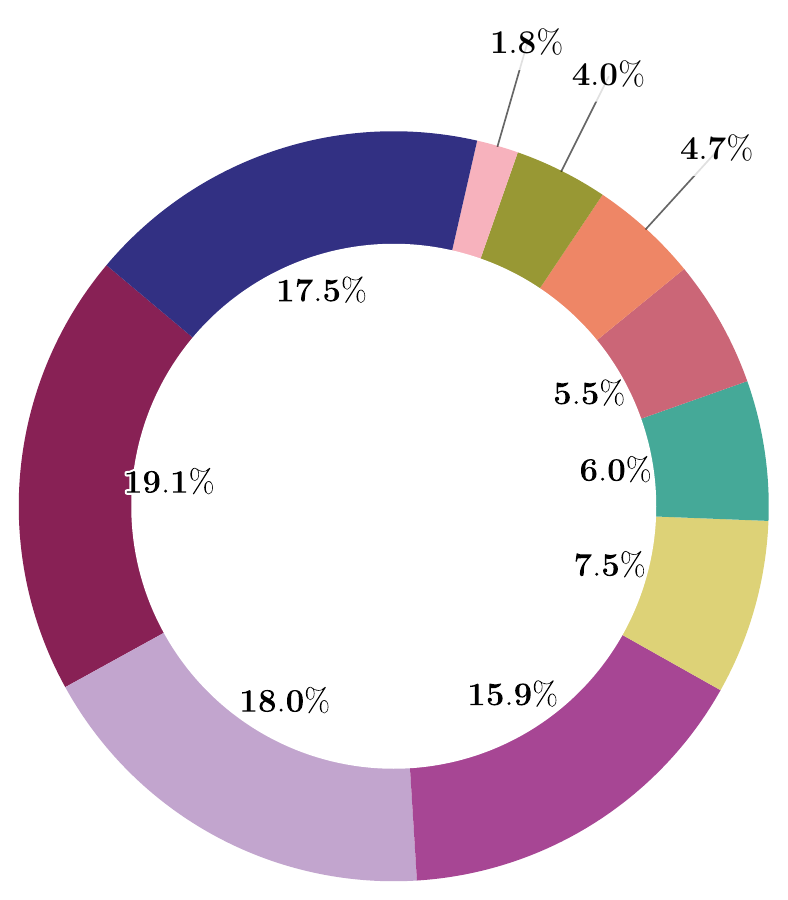}
\caption{Image (Non-Celebrity)}
\label{figure:non_celebrity_image_models_donut}
\end{subfigure}
\\[1ex]
\begin{subfigure}[b]{0.25\linewidth}
\centering
\includegraphics[width=\linewidth]{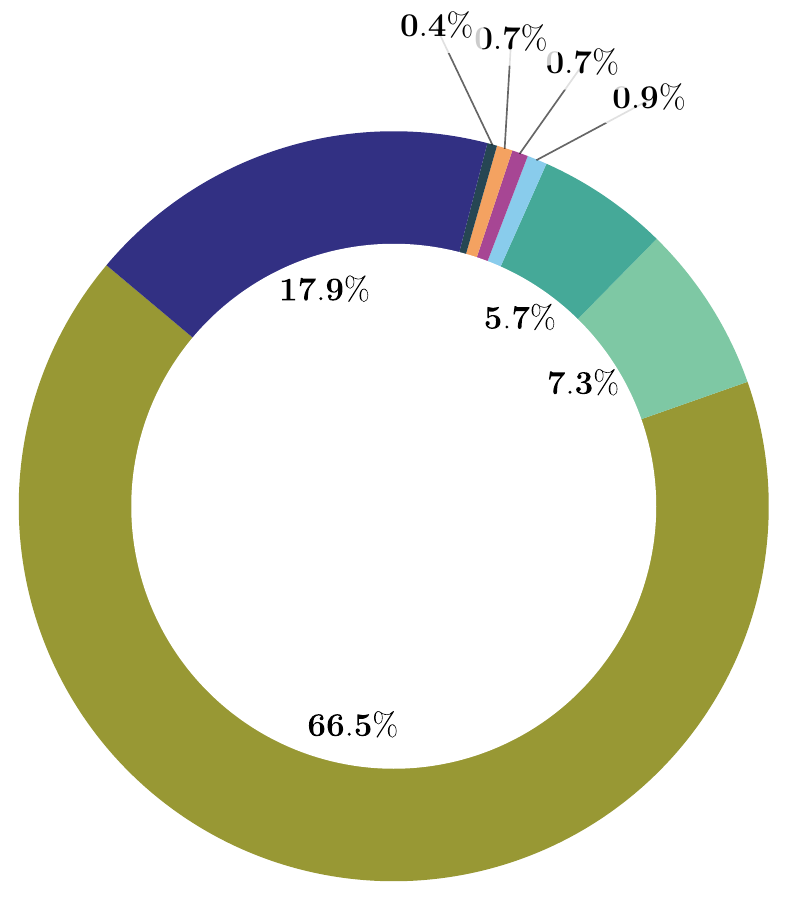}
\caption{Video (All Targets)}
\label{figure:all_video_models_donut}
\end{subfigure}
\hfill
\begin{subfigure}[b]{0.25\linewidth}
\centering
\includegraphics[width=\linewidth]{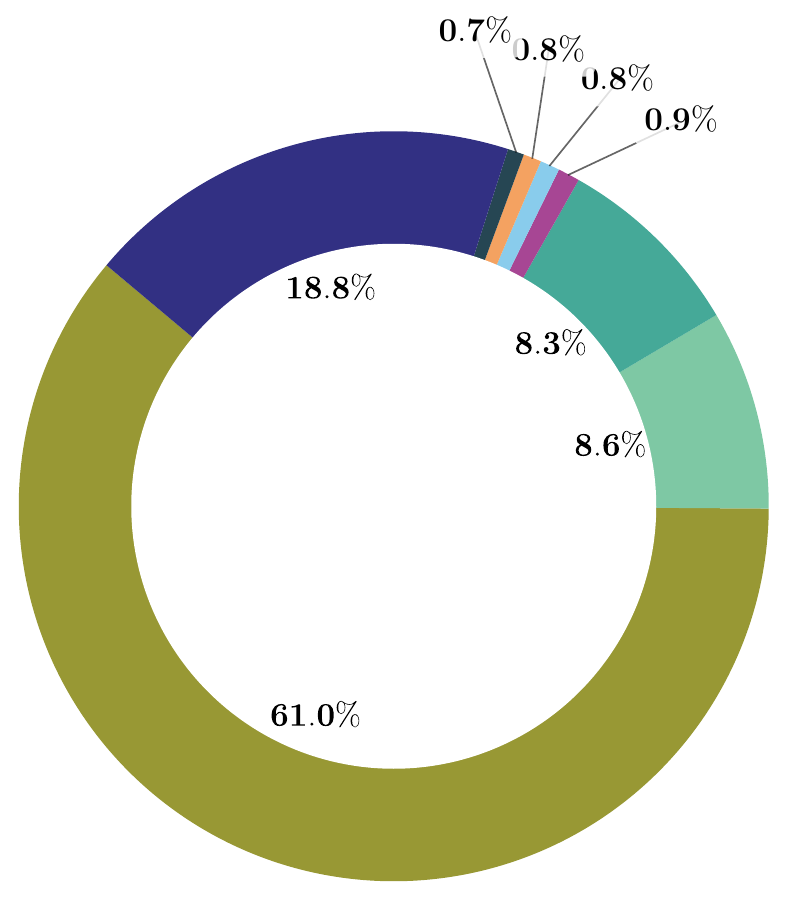}
\caption{Video (Celebrity)}
\label{figure:celebrity_video_models_donut}
\end{subfigure}
\hfill
\begin{subfigure}[b]{0.25\linewidth}
\centering
\includegraphics[width=\linewidth]{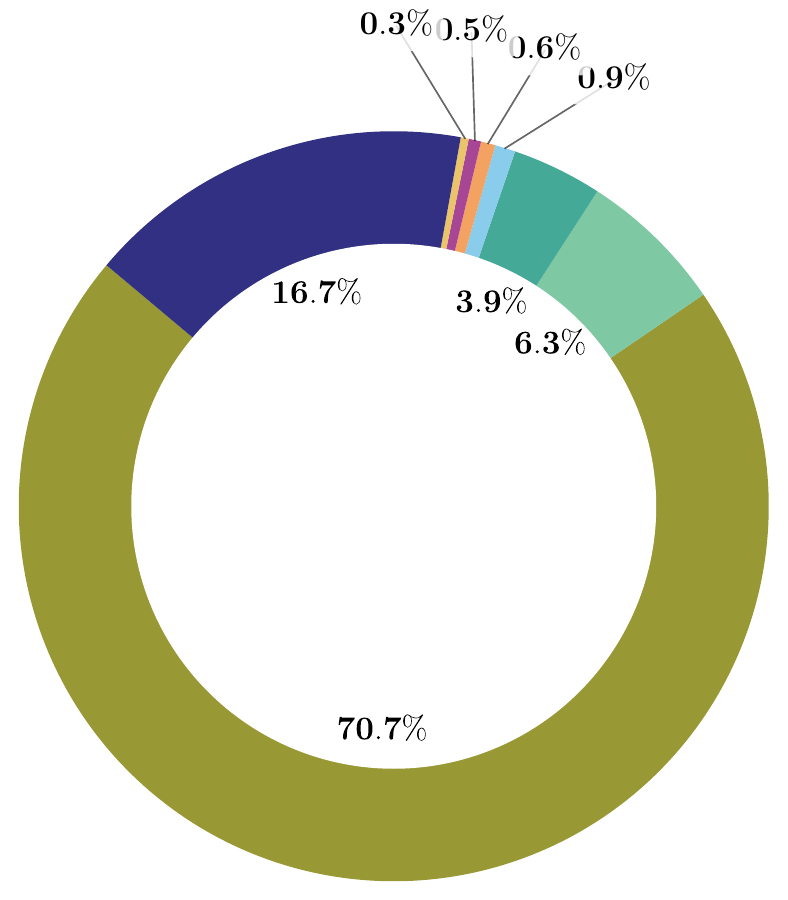}
\caption{Video (Non-Celebrity)}
\label{figure:non_celebrity_video_models_donut}
\end{subfigure}
\end{minipage}
\hspace{0.01\textwidth}
\begin{minipage}[c]{0.15\textwidth}
\centering
\includegraphics[width=\linewidth]{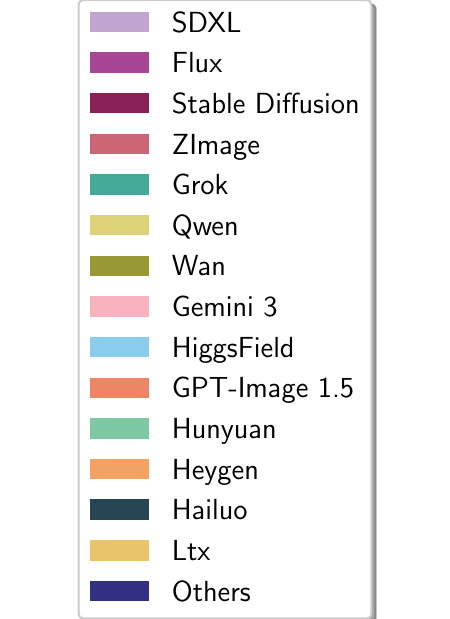}
\end{minipage}
\caption{Source model distribution of SNEACI, disaggregated by target type and media format.}
\label{figure:models_distribution_by_target}
\end{figure*}

\mypara{Overall Model Distribution}
As shown in~\autoref{figure:all_image_models_donut}, the Stable Diffusion family is the primary driver of image-based AI nudification.
SDXL~\cite{PELBDMPR23} and Stable Diffusion~\cite{RBLEO22} together account for 42.4\% of all scene synthesis images~\footnote{In this paper, we treat SDXL separately from Stable Diffusion because SDXL is widely used and is explicitly recognized as a distinct model in AI nudification.
Accordingly, \textit{Stable Diffusion} refers only to Stable Diffusion itself, whereas the \textit{Stable Diffusion family} includes both Stable Diffusion and SDXL.}; Flux~\cite{flux} follows at 15.3\%.
These findings align with previous research: Hawkins et al.~\cite{HMR25} found that Stable Diffusion family and Flux are the most commonly shared base models for deepfake image fine-tuning on public repositories (e.g., HuggingFace~\cite{huggingface}, Civitai~\cite{Civitai}).
For videos, as~\autoref{figure:all_video_models_donut} shows, Wan~\cite{WAWMXCYZYZWZZWCZZYHMZLWCFZSFWGWSLWWZWSYSHXKLLLWZHLWLPZHSFJHWL25} accounts for 66.5\% of all AI-generated videos, with Hunyuan~\cite{KTZMDZXLWZWLYLWWLHYTWSBWXWWLLLWYDLCCPYHXZXTLLZWYWLJZ24} a distant second at 7.3\%.

We observe that all of the predominant models (e.g., Stable Diffusion family and Wan) are open-source, which means their weights can be downloaded freely, fine-tuned on custom data, and invoked without any safety filtering.
Critically, these models lack effective guardrails against NSFW content generation, including sexual imagery.
Previous study from Schneider et al.~\cite{SH25} has shown that Stable Diffusion family can generate sexual images from straightforward prompts, though they are less effective at depicting specific named individuals without additional fine-tuning.
However, this limitation can be easily overcome through identity-specific LoRA fine-tuning, which we discuss later in~\autoref{section:rq2_shared_models}.
For videos, although the developers of Wan reports filtering all inappropriate content from the training data~\cite{WAWMXCYZYZWZZWCZZYHMZLWCFZSFWGWSLWWZWSYSHXKLLLWZHLWLPZHSFJHWL25}, the base model can still be fine-tuned to produce NSFW content~\cite{KCDYRBH25}.
In both cases, the open-source release model undermines centralized oversight, as no single entity can prevent or monitor downstream misuse once models are publicly released.
\finding{Open-source models lacking safety guardrails dominate the SNEACI ecosystem: the Stable Diffusion family powers 42.4\% of images and Wan accounts for 66.5\% of videos.}

\mypara{Model Distribution of Target Type}
We further disaggregate model usage by target type (celebrity vs.\ non-celebrity) to examine whether different models are associated with different targets.
As shown in~\autoref{figure:celebrity_image_models_donut} and~\autoref{figure:non_celebrity_image_models_donut}, for celebrity images, SDXL is the most prevalent model, accounting for 36.4\%, while for non-celebrity images, SDXL accounts for only 18.2\%.
This suggests that SDXL is more commonly used for celebrity targets than for non-celebrity targets.
The reason behind is that many SDXL-based fine-tuned models are specifically designed for the nudification of certain celebrities, which are widely shared on public repositories, which we will further discuss in~\autoref{section:rq2_shared_models}.

For video content, as shown in~\autoref{figure:celebrity_video_models_donut} and~\autoref{figure:non_celebrity_video_models_donut}, the difference between celebrity and non-celebrity targets is relatively small.
Wan remains the dominant model in both categories, accounting for 61.0\% of celebrity videos and 70.7\% of non-celebrity videos, suggesting that target type has a weaker association with model choice in video generation than in image generation.
This contrasts with image generation, where the availability of identity-specific LoRAs likely drives stronger associations between target type and model choice.
In video generation, the prevalence of image-to-video pipelines, where a single image of the target can be directly animated, reduces the need for identity-specific model training, narrowing the technical distinction between celebrity and non-celebrity targets.
\finding{The proliferation of target-specific fine-tuned models drives image production, while image-to-video pipelines democratize video production, making any individual with a single photograph a potential target.}

%-------------------------------------------------------------------------------
\subsection{Resources Shared in the Community}
\label{section:rq2_resources}
%-------------------------------------------------------------------------------

Beyond the generation models themselves, the AI nudification ecosystem relies on a supply chain of shared resources, including generation techniques, fine-tuned models, and various platforms.
These resources lower the technical and financial bars for SNEACI production.
By allowing non-experts to create such media easily, they lead to more widespread abuse and broader social impact.
To understand this supply chain, we analyze the resources explicitly shared in 4chan threads.
Because \textit{Adult Requests} board is oriented toward content exchange rather than technical discussion, in-depth workflow tutorials are rare.
Nonetheless, some threads still contain links to external model repositories and generation techniques, which we systematically collect and analyze below.

%-------------------------------------------------------------------------------
\subsubsection{Models}
\label{section:rq2_shared_models}
%-------------------------------------------------------------------------------

\begin{figure}
\centering
\includegraphics[width=0.8\columnwidth]{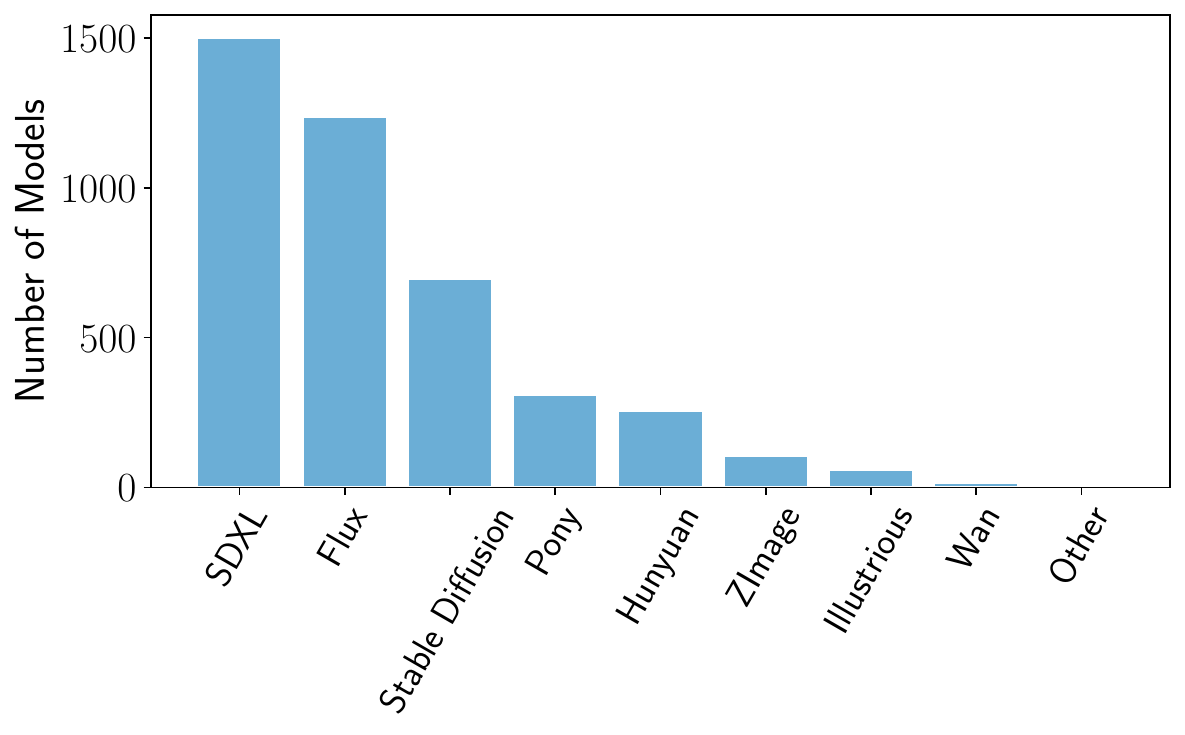}
\caption{Base model distribution of 4,216 fine-tuned models shared within the community, with SDXL (35.9\%) and Flux (29.5\%) as dominant architectures.}
\label{figure:shared_models_base_model}
\end{figure}

We first analyze models shared within the community, drawing from two sources: links explicitly posted in threads, and links referenced in technique guides.
In total, we collect 4,216 accessible model links, among them 1,451 are NSFW models, which are explicitly described as NSFW by their publishers.
We further categorize these NSFW models into the fine-tuning type.
We find 864 identity-specific LoRA fine-tuning, which accounts for 59.6\%, and the rest are general fine-tuning for a specific sence or a group of people with certain characteristics.

Furthermore, we use \texttt{gemini-3-flash-preview} to extract celebrity names and genders from model titles.
We find that 98.6\% of NSFW identity-specific LoRAs target female celebrities, involving 492 different targets.
This aligns with prior findings by Hawkins et al.~\cite{HMR25}, who report that deepfake models overwhelmingly target female celebrities.
As celebrities have abundant photographs available online, it is more feasible to fine-tune identity-specific models targeting them.
Moreover, their public recognition means that these models attract greater attention and circulate more widely, which further incentivizes the creation and sharing.

We also analyze the base model of the shared fine-tuned models.
As shown in~\autoref{figure:shared_models_base_model}, SDXL is the most common base model among shared LoRAs, accounting for 35.9\%, followed by Flux (29.5\%).
This distribution is similar to model provenance results in~\autoref{section:rq2_provenance}, confirming that the models shared in the community are used to produce SNEACI in the wild.

Notably, we find that 98.7\% of NSFW models are shared through the same publicly accessible platform,\footnote{This platform hosts a large number of policy-violating models for SNEACI production.
To the best of our knowledge, this platform has not been documented in prior work; we therefore withhold its name to avoid directing further attention to it.} which serves as an archive for models removed from other hosting platforms.
Among the models shared on this platform, 99.6\% were originally published on Hugging Face or Civitai, and 63.0\% have been deleted from the original platform at least once but remain downloadable through alternative links.
This suggests that once harmful models have been publicly released, removing them from the Internet becomes difficult in practice.
Accordingly, model sharing platforms should focus more on preventing harmful models from being published in the first place, rather than deletion after release.

\finding{NSFW fine-tuned models shared in the community overwhelmingly target female celebrities, are actively used to produce SNEACI in the wild, and remain difficult to permanently remove from the Internet.}

%-------------------------------------------------------------------------------
\subsubsection{Techniques}
\label{section:rq2_techniques}
%-------------------------------------------------------------------------------

Among the collected threads, we find two links to external guides that document end-to-end workflows for generating SNEACI: one for SDXL-based image generation and one for Wan-based video generation.
Despite targeting different modalities, the two guides share a common structure: both walk users from software installation through model selection, target-specific customization, and output refinement, forming a complete pipeline that requires no prior machine-learning expertise.

\mypara{Image Generation (SDXL)}
The image guide is organized around a three-stage pipeline: (1) install a graphical UI front-end (e.g., ComfyUI~\cite{comfyui}), (2) download SDXL checkpoint as the base model, and (3) load one or more LoRA weights (e.g., identity-specific LoRA).
When combined with a base model, they enable users to generate identity-preserving images of the target using a single prompt.
Besides the whole pipeline, this guide also provides links to the model repositories where users can find and download LoRAs for specific celebrities, as we discussed in~\autoref{section:rq2_shared_models}

\mypara{Video Generation (Wan)}
Unlike the image guide, which focuses on AI nudification, the video guide is framed as a general tutorial for using the Wan~2.1~\cite{wan2_1}, which is not specifically designed for nudification.
The video guide follows a similar pattern but centers on an image-to-video workflow: users supply a reference image of the target and the Wan~2.1 model generates a short clip.
And the guide uses a large portion to discuss how to optimize the generation process, reducing the inference time from 50 minutes to 5 minutes on consumer hardware.

Taken together, the two guides illustrate how the nudification ecosystem packages complex generative-AI pipelines into step-by-step recipes.
Additionally, even though some guides do not focus on generating sexual content, the same workflows and models can be easily applied to AI nudification.

%-------------------------------------------------------------------------------
\subsubsection{Platforms}
\label{section:rq2_platforms}
%-------------------------------------------------------------------------------

As a content-sharing community, 4chan serves as a primary venue for users to exchange SNEACI and related resources.
Beyond SNEACI posted directly on the board, we observe a large number of external links shared in threads.
We collect all URLs and, after filtering out links unrelated to AI nudification, classify the remaining into four categories based on their roles in the supply chain.
These categories follow the AI nudification pipeline from capability acquisition to content production, distribution, and off-platform migration.

\begin{itemize}
\item\textit{Resource-sharing platforms}.
As discussed in~\autoref{section:rq2_shared_models} and~\autoref{section:rq2_techniques}, these platforms primarily provide public access to models and techniques related to AI nudification.
They allow users to obtain the technical resources needed to carry out nudification, thereby serving as an important source of capability provision in the ecosystem.

\item\textit{Ready-to-use web applications}.
This category includes both general-purpose generative platforms (e.g., Grok~\cite{grok}, Meta AI~\cite{meta_ai}) and websites specifically designed for AI nudification~\cite{GOBCBTRK25}.\footnote{Even for general-purpose platforms with content restrictions, Mink et al.~\cite{MQR26} suggest that jailbreak techniques may still enable the generation of such content; we also observe such cases in the 4chan threads, although only in a small number of instances.} 
For services specifically designed for AI nudification, they substantially lower the technical barrier by allowing users to generate content directly through a web interface. 
Additionally, some links to such services are shared in ways that appear intended to attract new users, suggesting active promotion and diffusion of these platforms within the ecosystem.

\item\textit{Content-delivery platforms}.
These platforms are used to share and deliver already generated SNEACI (e.g., Gofile~\cite{gofile}, Catbox~\cite{catbox}), often through file-hosting links.
Compared with directly posting such content on the board, this form of distribution is more discreet and allows users to exchange abusive outputs in a less visible manner.
Compared with mainstream platforms, these services often adopt looser moderation of NSFW content~\cite{catbox_policy}, making them more conducive to the sharing of SNEACI.

\item\textit{Private community platforms}.
We further observe invitation links to private or semi-private communities, such as Discord and Telegram.
These links suggest that AI nudification-related activities are not confined to public threads, but may extend into more persistent and less visible spaces, facilitating off-platform user migration and more concealed forms of community organization.
\end{itemize}

\finding{External platforms linked within the community span four distinct roles, from capability provision and content production, to distribution and off-platform migration, collectively lowering barriers to entry while enabling more concealed forms of abuse.}

%-------------------------------------------------------------------------------
\section{RQ3: Characterizing the Ecosystem}
\label{sec:rq3}
%-------------------------------------------------------------------------------

Having characterized the content SNEACI ecosystem produces (RQ1) and the supply chain that enables production (RQ2), we now turn to the social dynamics that sustain it, examining how users interact within the request-fulfill ecosystem on 4chan.

%-------------------------------------------------------------------------------
\subsection{Interaction Model}
\label{section:rq3_interaction_model}
%-------------------------------------------------------------------------------

Beyond spontaneously sharing SNEACI within the community, one major mode of interaction is request and fulfillment between users.
We formalize this interaction by defining two primary roles:
\begin{itemize}
\item \textbf{Requesters} are users who seek SNEACI but generally lack the technical expertise to produce it themselves.
They submit SNEACI requests that typically contain two elements: \textit{request text} specifying the desired output, such as ``nudify,'' ``undress,'' or a particular sexual scenario; and a \textit{request image},  a source photograph of the target individual which is usually SFW.
Although a request often includes both elements, either one can also appear alone.
For example, a request may explicitly name the target when the target is well known, while in specific topic threads, a single target photograph can also function as a request.
\item \textbf{Providers} are users who possess the generative-AI capabilities (e.g., models, hardware, and workflow knowledge), which are needed to fulfill these requests.
They respond by posting SNEACI depicting the requested target.
And they are often referred to as ``Wizards'' in the community.
\end{itemize}

\begin{figure}
\centering
\includegraphics[width=0.8\columnwidth]{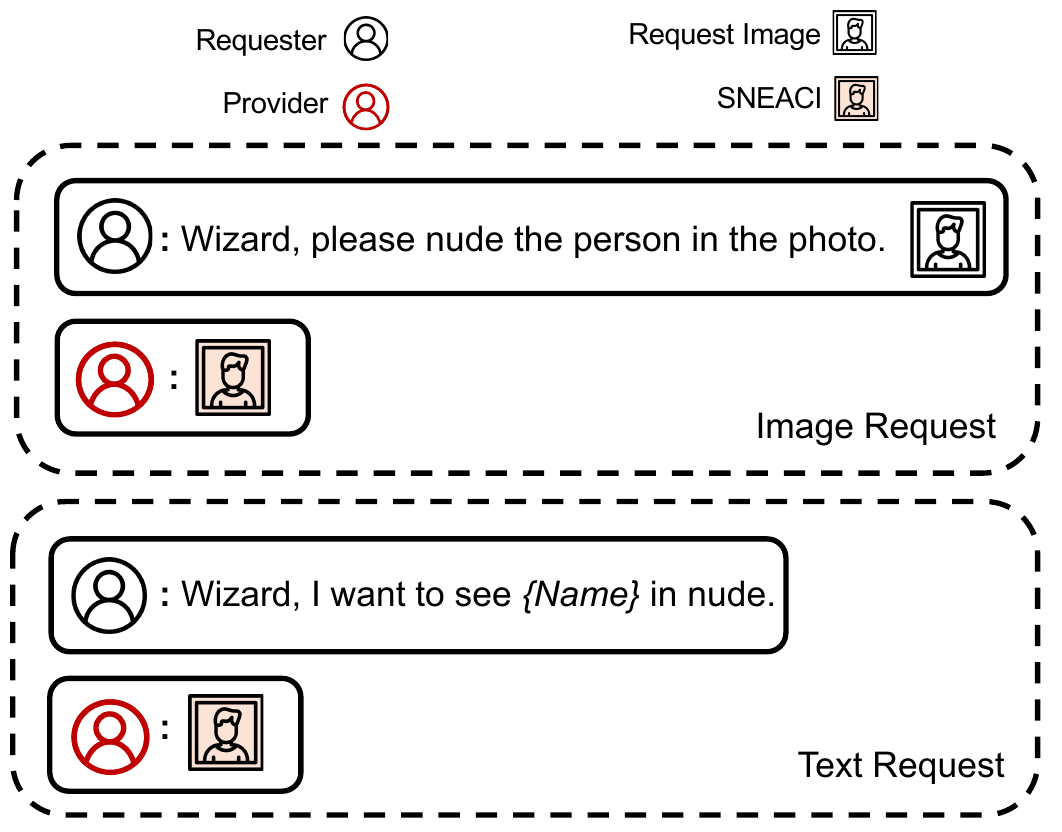}
\caption{Examples of SNEACI request-fulfill interactions, showing image requests and text-only requests.}
\label{figure:request_response}
\end{figure}

As shown in~\autoref{figure:request_response}, a typical exchange proceeds as follows: a requester creates a thread or posts within an existing one, attaching a source photograph and a textual prompt describing the desired output.
If a provider is active in the thread, they select the request, generate the content, and post it as a reply, sometimes tagging the original requester.
In some cases, providers take a more proactive role by creating dedicated threads that advertise a ``menu'' of available generation services, establishing explicit rules for participation.
The following excerpt from one such thread illustrates this format:

\begin{quote}
\textit{... Here are the rules:}\\
\textit{1.\ Post one picture}\\
\textit{2.\ Select one of the 3 effects}\\
\textit{3.\ Wait for me to do it}\\
\textit{4.\ When you receive your video from me, you can then request another one.}\\
\textit{Here are the 3 effects you can choose from:}\\
\textit{Sexual Scene 1 / Sexual Scene 2 / Sexual Scene 3 ...''}
\end{quote}

%-------------------------------------------------------------------------------
\subsection{Data Processing}
\label{section:rq3_data_processing}
%-------------------------------------------------------------------------------

To investigate the production and consumption of SNEACI, we analyze the posts collected in~\autoref{section:rq1_data_collection}, extract the requests, and identify the corresponding responses for each request.
Furthermore, we examine the target types of SNEACI requests (celebrity vs.\ non-celebrity).

\begin{figure}
\centering
\includegraphics[width=\linewidth]{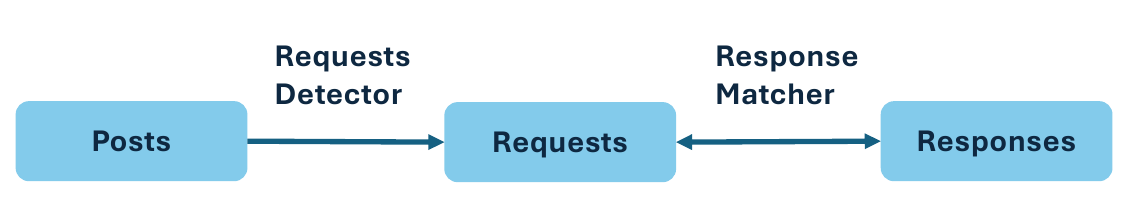}
\caption{Pipeline for extracting and matching SNEACI requests and responses from 4chan posts using request detector and response matcher.}
\label{figure:pipeline2}
\end{figure}

\mypara{Overview}
The overall data processing pipeline is illustrated in~\autoref{figure:pipeline2}.
First, we iterate through all collected posts and employ \textit{Request Detector} to identify posts soliciting SNEACI.
Second, for each identified request, we search for its corresponding response within the same thread by using the \textit{Response Matcher}.
Finally, we classify the target of each request as either a celebrity or non-celebrity using the \textit{Request Target Classifier}.

\mypara{Classifiers and Detectors}
We use the following classifiers and detectors in our pipeline, and details of their implementation are provided in~\autoref{appendix:implementation_details_of_detectors}.

\begin{itemize}
\item \textbf{Request Detector.}
We design a classifier to identify if a post is a request for SNEACI.
For each post, we first analyze the request image, if any, using our NSFW detector in~\autoref{section:rq1_data_processing} to determine whether the image is SFW.
We then provide this image-level SFW label together with the post text to \texttt{GPT-4o-mini}~\cite{GPT-4o-mini}, which judges whether the post expresses a request for SNEACI.
We manually verify the request detector on 200 randomly sampled posts and find that it achieves 95.0\% accuracy.

\item \textbf{Response Matcher.}
A request typically either receives a response in the same thread or receives no response at all.
To identify the response of each request, we examine all SNEACI items posted in that thread and determine whether any of them depicts the same person as the target in the request image.
Because InsightFace has been widely adopted in prior work on face recognition~\cite{LFSWHZPZS25}, we use it for face matching by computing the similarity between the face in the request image and that in a candidate SNEACI item.
If at least one SNEACI item in the same thread exceeds a predefined facial similarity threshold, we consider the request responded to; otherwise, we consider it unresponded.
For requests without request images, we extract the named entities from the request text and check if any SNEACI in the thread is targeting the same person.

\item \textbf{Request Target Classifier.}
To further analyze each request's target, we classify whether the request is for a celebrity or a non-celebrity.
We first check if text contains a celebrity name directly; if not, we apply the celebrity classifier in~\autoref{section:rq1_data_processing} to the request image to determine whether the target is a celebrity or non-celebrity.
\end{itemize}

%-------------------------------------------------------------------------------
\subsection{Results}
%-------------------------------------------------------------------------------

\begin{table}[!t]
\centering
\caption{Request and response statistics by target type, showing 74.75\% of requests target non-celebrities with overall response rate of 22.60\%.}
\label{table:requests_responses}
\scalebox{0.7}{
\begin{tabular}{l | c | c | c}
\toprule
& Celebrity & Non-Celebrity & Total\\
\midrule
Requests Count & 4,261 &  12,615 & 16,876\\
Responses Count & 1,074  &  2,740 & 3,814\\
\midrule
Response Rate & 25.21\% & 21.72\% & 22.60\%\\
\bottomrule
\end{tabular}}
\end{table}

%-------------------------------------------------------------------------------
\subsubsection{Requests and Responses}
%-------------------------------------------------------------------------------

We identify a total of 16,876 requests, which account for 21.00\% of all posts.
We then classify the requests based on their targets.

\mypara{Targets of Requests}
As shown in~\autoref{table:requests_responses}, 25.25\% of requests target celebrities, while 74.75\% target non-celebrities.
This helps explain why non-celebrity-targeted content is more prevalent in the media analysis in~\autoref{section:rq1_targets}: the demand for non-celebrity targets is substantially higher than that for celebrity targets.
This finding aligns with our broader argument that, as technical barriers decline, AI nudification becomes increasingly accessible.
As a result, AI nudification is shifting from the targeting of celebrities, which has received greater public attention, toward the targeting of personal acquaintances, making this form of abuse more intimate and socially proximate.

\mypara{Response Rates}
We next examine the rate at which requests receive responses.
Overall, 22.60\% of all requests receive a matched response from providers.
As shown in~\autoref{table:requests_responses}, celebrity requests exhibit a slightly higher response rate (25.21\%) compared to non-celebrity requests (21.72\%).
However, despite the higher per-request response rate for celebrities, non-celebrity responses outnumber celebrity responses by more than twofold.
That is because the number of non-celebrity requests is substantially larger than that of celebrity requests (12,615 vs.\ 4,261).
This suggests that when given the opportunity to request AI nudification, users predominantly target people within their own social circles rather than public figures.

\finding{Non-celebrity requests dominate the board and account for more than twice the absolute volume of fulfilled content than celebrity requests, confirming that demand drives the prevalence of non-celebrity targets.}

%-------------------------------------------------------------------------------
\subsubsection{Active Providers}
\label{section:rq3_active_suppliers}
%-------------------------------------------------------------------------------

\begin{figure}
\centering
\begin{subfigure}{\columnwidth}
\centering
\includegraphics[width=0.8\columnwidth]{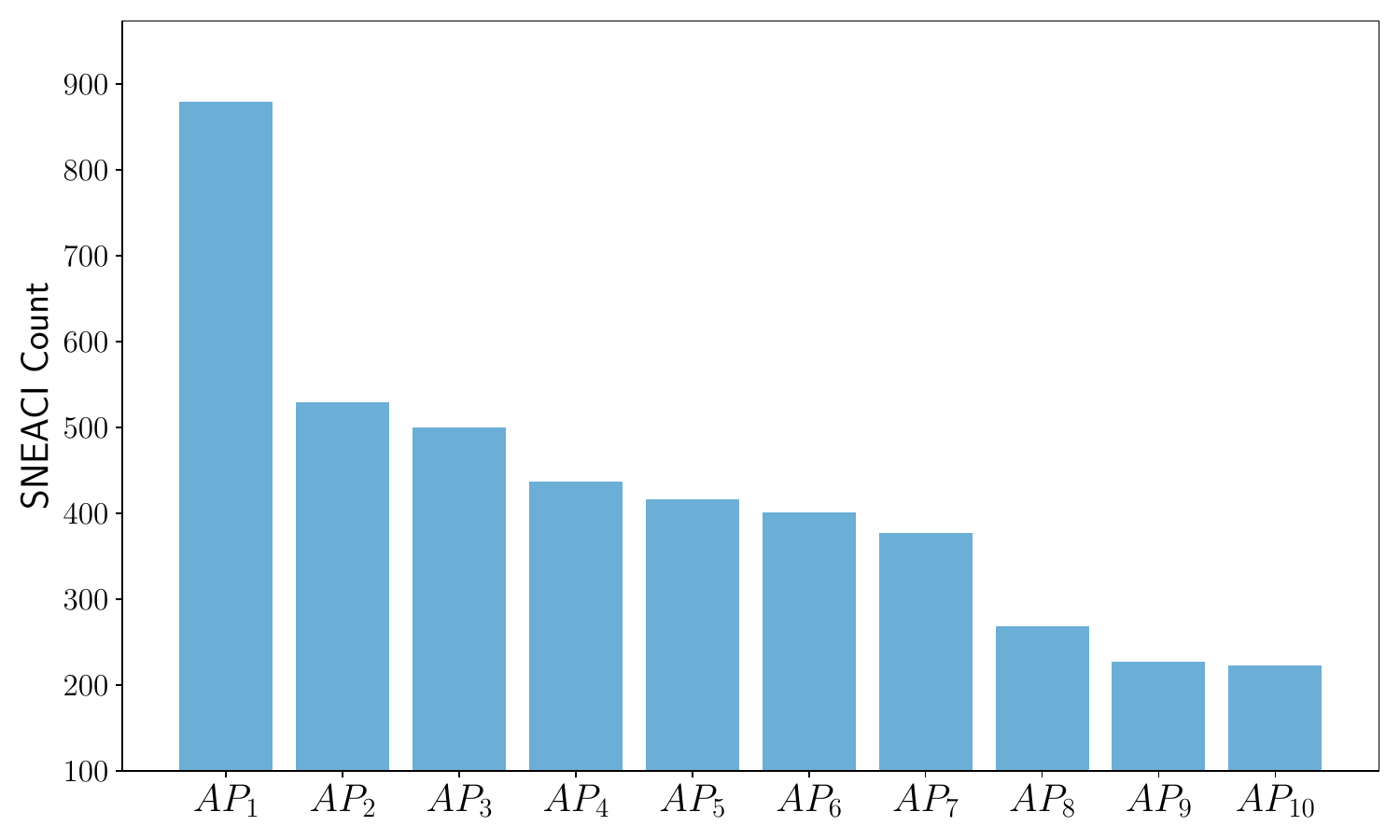}
\caption{SNEACI count of top 10 active providers}
\label{figure:author_media_count}
\end{subfigure}
\begin{subfigure}{\columnwidth}
\centering
\includegraphics[width=0.8\columnwidth]{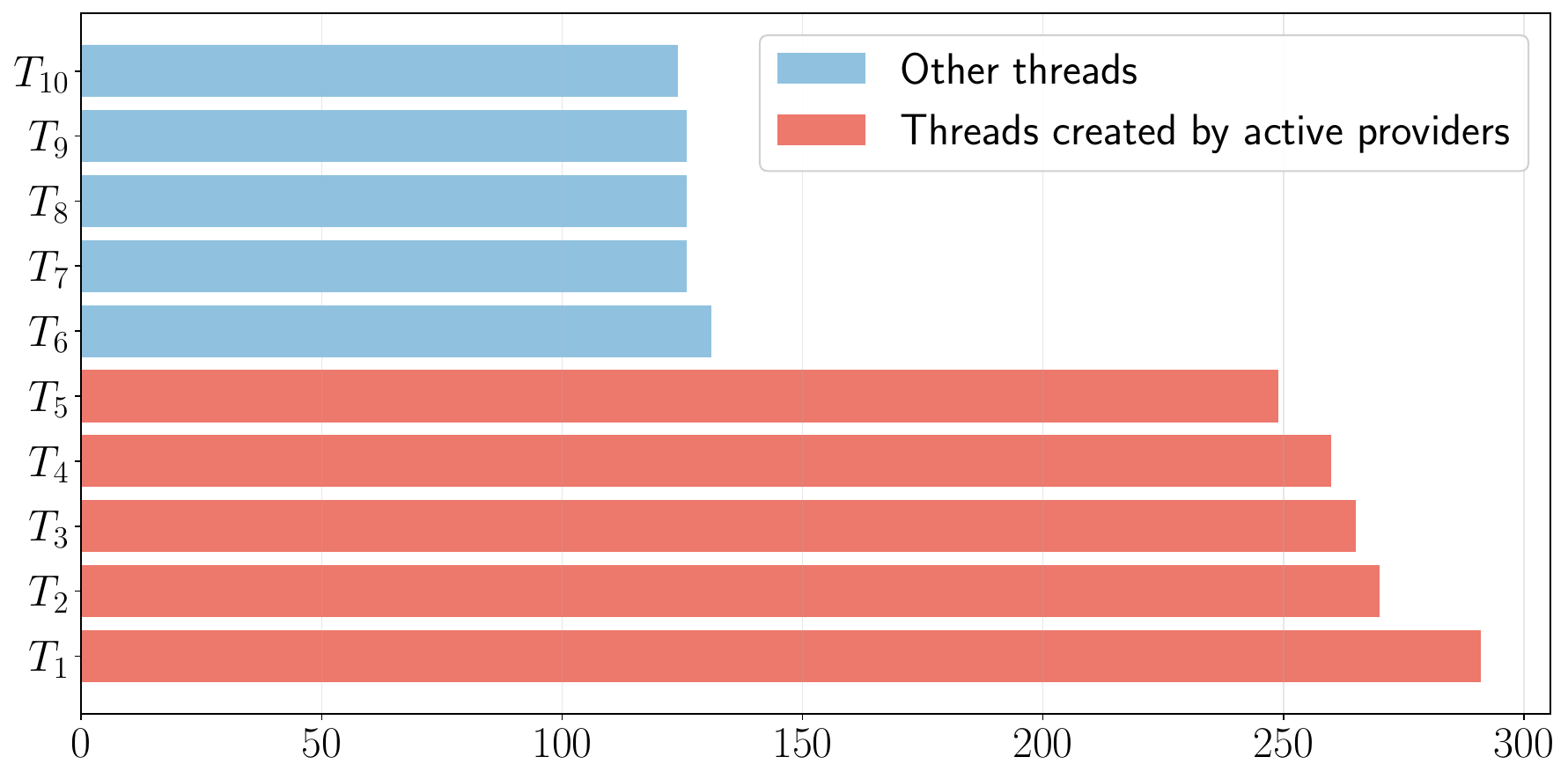}
\caption{Request volume in top 10 threads}
\label{figure:requests_by_thread}
\end{subfigure}
\caption{Distribution of SNEACI production across active providers and request volume across threads.
($AP$ denotes the active provider, $T$ denotes the thread.)}
\label{figure:active_suppliers_influence}
\end{figure}

\begin{figure*}
\centering
\includegraphics[width=0.8\textwidth]{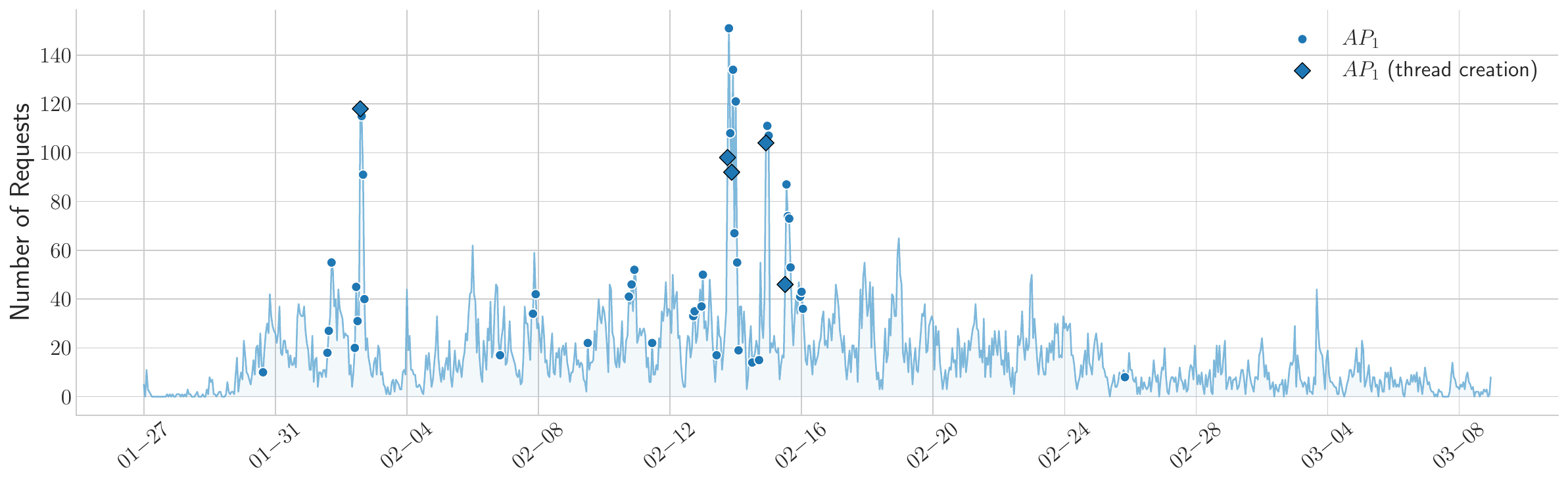}
\caption{Request volume over time during January 27 - March 08, 2026, with spikes aligned to $AP_1$ thread creation.}
\label{figure:requests_over_time}
\end{figure*}

Having characterized the types of requests and their response rates, we next examine the providers who actively fulfill these requests.
Although 4chan allows anonymous posting, and 90.96\% of posts on the \textit{Adult Requests} board are made by anonymous users, we observe a small amount of non-anonymous users\footnote{On 4chan, users named \textit{Anonymous} are anonymous; other usernames are treated as non-anonymous.} consistently produce SNEACI, making them a distinctive and analytically tractable subset of the community.
We refer to them as \textit{active providers} and further analyze their proportion, motivation and community-level influence.

\mypara{Identification}
We define an active provider as a non-anonymous user who posted at least five SNEACI items during the observation period.
Through this definition, we capture 61 active providers.
These 61 users collectively posted 4,981 SNEACI items, accounting for 20.66\% of all SNEACI in our dataset.
As shown in~\autoref{figure:author_media_count}, the distribution is heavily skewed: provider generating the most SNEACI content (i.e., $AP_1$) alone contributed 780 items, and the top 10 providers account for 65.45\% of the 4,981 items.
Despite representing a small fraction of the user base, these active providers thus produce a disproportionate share of all SNEACI on the board.

\mypara{Motivation}
These active providers appear to be motivated by at least two factors.
First, community recognition.
We find that some active providers gain visibility and reputation by frequently responding to the requests.
Their presence can in turn attract more requesters to post requests, a pattern that we analyze in more detail below.
Second, financial gain.
Some providers appear to use the free content they post on 4chan as a portfolio to attract paying customers for customized nudification services on private platforms.
They would provide their contact information in the thread or in their names, such as \texttt{\{username\}@discord} or \texttt{telegram:\{username\}}, to offer private commission services, which typically involve a fee.

\mypara{Active Providers as Ecosystem Architects}
Active providers do not merely respond to existing demand; they actively catalyze and shape it.
This influence operates through three mechanisms: catalyzing request volume, exploiting requesters to amplify harm, and cultivating new providers.

\begin{itemize}
\item \textbf{Catalyzing Request Volume.}
The presence of an active provider substantially increases the number of requests submitted by other users.
As shown in~\autoref{figure:requests_by_thread}, among the top 10 threads ranked by request count, the top 5 are all created by active providers and these threads exhibit significantly higher request volumes.
The temporal analysis in~\autoref{figure:requests_over_time} further reveals the influence of active providers on request volume.
Requests peak when $AP_1$ creates new threads, demonstrating that $AP_1$'s visible provider activity directly drives request volume.
The visible presence of a responsive provider incentivizes requesters to submit requests, creating a feedback loop between provider activity and request volume.

\item \textbf{Exploiting Requesters to Amplify Harm.}
Beyond stimulating volume, active providers also shape \textit{what} is requested, and in doing so, deepen the harm inflicted on targets.
As described in~\autoref{section:rq3_interaction_model}, some active providers create dedicated solicitation threads that impose explicit constraints on acceptable requests, effectively acting as gatekeepers who filter the pool of potential targets.
Among the top 10 active providers, 6 have created at least one such thread.
These constraints operate at multiple levels.
Some providers restrict requests to non-celebrity targets, reinforcing the shift toward targeting people in requesters' own social circles.
For example, one provider's thread states:

\begin{quote}
\textit{... - RULES: Nothing that breaks site rules.} \\
\textit{- Must be Asian.} \\
\textit{- IRLs\footnote{IRLs means people in real life, which are non-celebrities.} only.} \\
\textit{- no celebs / ig / etc ...}
\end{quote}

\noindent Others impose demographic preferences, narrowing acceptable targets by race, gender, age, or body type.
The rule above explicitly requires Asian targets; another provider states a preference for ``pretty, young, women,'' illustrating how provider preferences filter the pool of potential targets:

\begin{quote}
\textit{... send me who they are to you and their name for higher chances >;3 im nosy like that.I prefer pretty, young, women ...}
\end{quote}

\noindent Most concerningly, some providers require requesters to disclose personal information about the target, including their name and relationship to the requester, compounding the harm by linking generated content to real-world identities and drawing requesters into deeper complicity.
For example, it states: 

\begin{quote}
\textit{... Tell me who the girl is to you and her name. the more info the more likely i am to choose her, im sentimental like that...}
\end{quote}

\noindent By setting these rules, providers act as gatekeepers who shape the distribution of harm.
They determine not only whether a request is fulfilled but also which categories of targets are deemed acceptable, thereby shaping the distribution of harm.

\item \textbf{Cultivating New Providers.}
Beyond producing content and directing requests, active providers also disseminate technical knowledge within the community.
Among the 61 active providers, 17 share practical generation guidance in threads, including hardware recommendations, prompt engineering techniques, and LoRA training workflows.
Rather than merely lowering technical barriers, this knowledge transfer actively reproduces the ecosystem by expanding the pool of users capable of generating SNEACI independently, ensuring continuity even in the absence of the original providers.
\end{itemize}
Taken together, these three mechanisms reveal that active providers function not merely as content producers, but as ecosystem architects: they catalyze demand, exploit requesters to amplify harm against targets, and cultivate new providers, ensuring the ecosystem's self-perpetuation.

\finding{
Driven by community recognition and financial gain, a cohort of 61 active providers functions as ecosystem architects who extend far beyond content production: they catalyze request volume, exploit requesters to amplify harm against targets, and cultivate new providers, ensuring the self-perpetuation of the ecosystem.
}

%-------------------------------------------------------------------------------
\section{Discussion}
\label{section:discussion}
%-------------------------------------------------------------------------------

Our findings suggest that mitigating AI nudification requires interventions across the full ecosystem, from model development and distribution to content moderation and legal protection.

\mypara{Content Platforms}
One important reason why AI-generated abusive content is prevalent on 4chan is its anonymity and weak content restrictions.
Anonymity provides users with a sense of protection, enabling them to harm others with a reduced fear of accountability.
At the same time, the lack of meaningful content moderation creates a moral vacuum within the community, where the legitimacy of such behavior is rarely questioned.
In such an environment, AI nudification is implicitly normalized, lowering the psychological barrier to making abusive requests.
Therefore, a reasonable platform intervention is essential.
Moderation should not focus only on the SNEACI itself, but also on the surrounding infrastructure, including requests, technical guides, model-sharing links, and active providers.
Stronger enforcement against both harmful content and the coordination behaviors that sustain it may help disrupt the feedback loop between request and fulfillment.

\mypara{Open-Source Model Developers}
In recent years, AI capabilities in image and video generation have advanced rapidly.
The generated outputs have become increasingly realistic, while the cost of using these models continues to decline.
However, corresponding safeguards have lagged behind.
Our findings show that the two main open-source models used for AI nudification, Stable Diffusion family and Wan, both lack more effective protections.
Even though Wan claims to apply NSFW filtering to its training data, adapting it to generate NSFW content remains relatively easy.
We therefore encourage model developers to incorporate stronger safeguards before releasing models, such as more rigorous safety evaluations and red-teaming against harmful adaptation.
Even if such measures cannot fully prevent harmful repurposing, they can still raise the barrier to generating abusive content.

\mypara{Model Hosting Platforms}
Prior work~\cite{HMR25} has measured the prevalence of AI nudification models on hosting platforms in depth, and our findings further show how harmful models from these platforms play an important role in enabling AI nudification.
As discussed in~\autoref{section:rq2_shared_models}, we find that once a harmful model is released, it is difficult to remove it from the Internet entirely, even if the original platform later takes it down, 
Therefore, platforms should ensure that uploaded models do not explicitly promote abusive or unlawful uses, which should be reflected in their publication policies and review processes.
In~\cite{HMR25}, the authors also propose several practical mitigation strategies.
For example, analyzing model metadata without directly interacting with the model can be an effective way to identify suspicious uploads.
They also suggest using watermarking to trace harmful outputs back to the publishing account, as well as imposing stronger penalties, such as banning accounts that upload harmful models.

\mypara{Legal Protections and Public Awareness}
Safeguards deployed by developers and platforms alone are far from sufficient.
Addressing this issue effectively also requires the introduction of dedicated laws and policies, together with greater public awareness.
As AI nudification increasingly affects a wider population, virtually anyone whose photos can be accessed by others may become a potential target of abuse.
This broad and escalating risk makes legislative action an urgent necessity.
Encouragingly, some jurisdictions have already begun to respond through legislation.
For instance, the United States has introduced the \textit{Take It Down Act}~\cite{TAKE_IT_DOWN_ACT}, while relevant legislation in the United Kingdom has already taken effect~\cite{DEEPFAKE_LAW_UK}.
We further urge other countries or regions to accelerate legislative efforts in this domain, with particular attention to affected individual protection, platform responsibility, and effective mechanisms for content removal and redress.
Meanwhile, public awareness must also be enhanced so that people recognize that this is not merely a virtual act on the internet, but behavior that causes real harm~\cite{AS26} and must be met with corresponding responsibility.

%-------------------------------------------------------------------------------
\section{Limitation and Further Work}
%-------------------------------------------------------------------------------

\mypara{Scope and Time Window}
Our measurement is limited in both platform coverage and observation period.
We focus on a public community and therefore cannot observe SNEACI generated through end-to-end nudification services or shared in private communities.
Additionally, although our dataset covers only 41 days, it still contains 24,105 items.
Nevertheless, our findings reflect AI nudification activity in the community only during the time of collection, and behavior outside this window is not captured.
We encourage future SNEACI measurement work to analyze longer-term temporal trends, identify key stages in community evolution, and uncover the factors shaping these changes.

\mypara{Classifier Accuracy}
Our analysis depends on automated classifiers and detectors with imperfect accuracy.
This is largely due to the complexity of the scenarios in our analysis, as current techniques are still unable to achieve substantially higher accuracy in this scenario.
For instance, our response matcher relies on face recognition, making it difficult to identify SNEACI with limited or unclear facial features, so the true response rate is likely higher than our estimate.

\mypara{Undressing Attribution}
In the model provenance analysis (\autoref{section:rq2_provenance}), we exclude undress image because current AIGC detectors have limited ability to identify such AI-edited images.
As detection methods improve, future work may offer a more complete understanding of the models used to produce this content.

\mypara{Shared Model Scope}
In shared models analysis (\autoref{section:rq2_shared_models}), we analyze only the models shared on 4chan.
Model-sharing platforms likely contain many additional AI Nudification models, but examining them is outside the scope of this paper, which focuses on AI nudification activity within 4chan \textit{Adult Requests}.
Future work could further study the broader technical ecosystem behind AI nudification.

%-------------------------------------------------------------------------------
\section{Conclusion}
%-------------------------------------------------------------------------------

We present a large-scale empirical study of AI nudification in the wild, analyzing 24,105 items of SNEACI from 4chan's \textit{Adult Requests} board over 41 days.
Our findings reveal that non-celebrity individuals are now the primary targets, indicating that this form of abuse is increasingly extending into everyday social relationships.
Open-source models with inadequate safety guardrails dominate production, supported by a large number of fine-tuned models and accessible tutorials that lower barriers to entry.
Requests and responses are the main interaction pattern, with 22.60\% of requests receiving responses.
Additionally, a small group of active providers further amplifies this dynamic.
These findings highlight the urgent need for coordinated interventions across multiple domains to disrupt the mechanisms sustaining this ecosystem.
We hope this work catalyzes collective action to establish more robust safeguards for affected individuals and curb the proliferation of AI-enabled image-based abuse.

\clearpage

%-------------------------------------------------------------------------------
\section*{Ethical Considerations}
%-------------------------------------------------------------------------------

This research involves the study of harmful non-consensual sexually explicit content, raising significant ethical considerations that we addressed through multiple safeguards.

\mypara{Platform Disclosure}
Mentioning 4chan’s \textit{Adult Requests} board may drive additional traffic to the platform.
We therefore considered whether to anonymize the platform.
However, because prior studies~\cite{CR26} and journalistic reports~\cite{4CHAN_DEEPFAKE_REPORT_JARED, 4CHAN_DEEPFAKE_REPORT_NIAMH} have already explicitly identified the platform and documented its conditions, we do not anonymize it in this paper.
By naming the platform, we aim to support more concrete policy discussion and intervention around SNEACI.

\mypara{Privacy Protection and Anonymization}
We take measures to protect the privacy of individuals who directly or indirectly appear in our dataset.
We mention the Taylor Swift incident in~\autoref{section:intro} because it has been widely covered in public reporting~\cite{Taylor_deepfake, DEEPFAKE_TAYLOR_SWIFT_CBS}, and discussed in prior research~\cite{GOBCBTRK25}.
Beyond this contextual reference, we do not mention any identifiable names in the paper.
Although some active provider used non-anonymous usernames on 4chan, we replace their usernames with anonymous identifiers (e.g., $AP_1$).
We make no attempt to track or identify these individuals beyond their on-platform activity.

\mypara{Data Collection}
4chan is a publicly accessible platform, and we only collected content that was openly available without any form of authentication or access restriction.
All research data has been restricted to the research team and has not been shared with any external parties, ensuring that affected individuals are not subjected to additional harm and that other researchers are not unnecessarily exposed to harmful content.

\mypara{Researcher Wellbeing}
Researchers involved in this study were informed in advance that they would be exposed to sexually explicit content as part of the data collection and analysis process.
To mitigate potential psychological harm, researchers have access to mental health resources throughout the study period.
And any researcher who feels distressed at any point can seek support and request a break without any negative consequences.

%-------------------------------------------------------------------------------
\section*{Open Science}
\label{section:open_science}
%-------------------------------------------------------------------------------

We release all data processing and detection pipeline code upon publication, shared in \url{https://anonymous.4open.science/r/AI_Nudification_Anon-E1F7/}.
This includes implementations of all the classifiers and detectors we used in our analysis, which are mentioned in~\autoref{section:rq1_data_processing} and~\autoref{section:rq3_data_processing}.
Due to the sensitive nature of the content and privacy concerns, we cannot release the raw dataset of SNEACI media files.
The dataset contains non-consensual sexually explicit imagery of real individuals, and public release would further harm the affected individuals and potentially violate privacy laws.
However, researchers who wish to verify our findings or conduct related studies may contact us to access anonymized metadata, including temporal distributions, target type labels, model attribution results, and aggregate statistics.

%-------------------------------------------------------------------------------
\section*{Generative AI Usage}
%-------------------------------------------------------------------------------

We used \texttt{GPT-5.5} for grammar checking and manuscript polishing.
We reviewed all generated content and take full responsibility for it.

%-------------------------------------------------------------------------------
\bibliographystyle{plain}
\bibliography{normal_generated_py3}
%-------------------------------------------------------------------------------

\appendix

%-------------------------------------------------------------------------------
\section{Durations of Video Content}
\label{appendix:durations_of_video_content}
%-------------------------------------------------------------------------------

\begin{figure}
\centering
\includegraphics[width=0.8\linewidth]{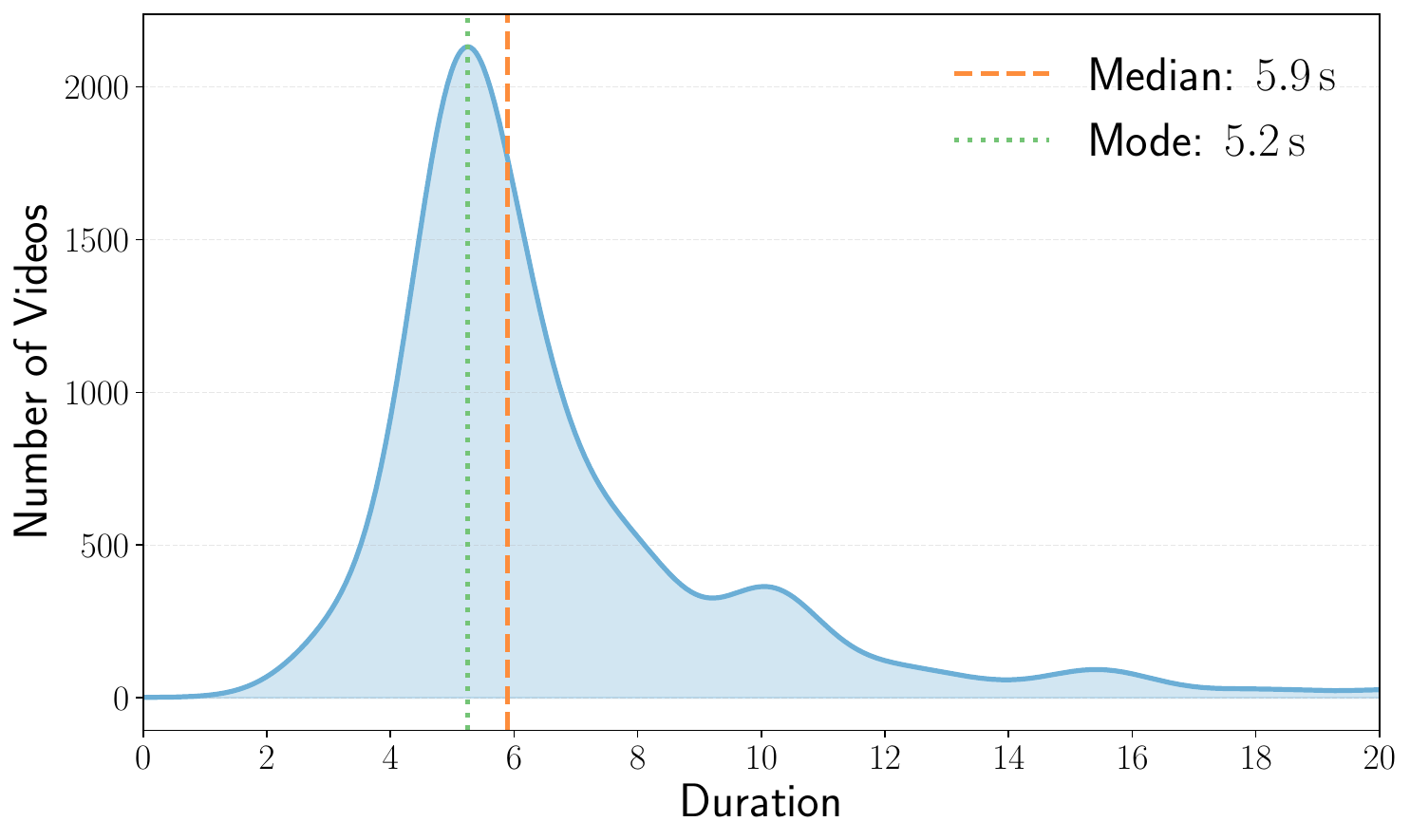}
\caption{Distribution of video durations across all SNEACI videos, showing heavy concentration around 5-8 seconds.}
\label{figure:video_duration_distribution}
\end{figure}

We analyze the duration distribution of 8,203 SNEACI videos.
As shown in~\autoref{figure:video_duration_distribution}, the distribution is heavily concentrated toward short durations.
In particular, 75\% of videos fall between 5.06 s and 8.06 s, while only 7.3\% exceed 15 s.
Additionally, the median duration is 5.90 s, and the most common duration is 5.20 s.

The concentration of videos in the 5.06-8.06 s range is consistent with the default output lengths of popular video generation models.
For example, the official Wan2.1~\cite{wan2_1} inference examples generate 81-frame videos at 16 fps, corresponding to a nominal duration of 5.06 s.
The prevalence of short-form content may also reflect practical constraints, as shorter videos are both cheaper and faster to generate.
For instance, with Wan2.1, a 5 s video can be generated on a single RTX 4090 in approximately 4 minutes, while generating longer videos requires disproportionately greater time and computational resources~\cite{WAWMXCYZYZWZZWCZZYHMZLWCFZSFWGWSLWWZWSYSHXKLLLWZHLWLPZHSFJHWL25}.

%-------------------------------------------------------------------------------
\section{Implementation Details of Detectors}
\label{appendix:implementation_details_of_detectors}
%-------------------------------------------------------------------------------

%-------------------------------------------------------------------------------
\subsection{Detectors and Classifiers Used in RQ1}
%-------------------------------------------------------------------------------

\mypara{NSFW Detector}
We use \texttt{Falconsai/nsfw\_image\_detection}~\cite{nsfw_detector} as the NSFW detector.
It is a fine-tuned Vision Transformer~\cite{DBKWZUDMHGUH21} on a proprietary dataset of approximately 80,000 images for binary classification (\texttt{nsfw} vs.\ \texttt{normal}).

\mypara{AIGC Detector}
We use the \texttt{AI-Generated \& Deepfake Content Detection} model provided by Hive AI.
Given an image or video as input, the model returns (1) the confidence scores for the input being real or AI-generated, and (2) if it is AI-generated, confidence scores for candidate source models.
We classify an input as AI-generated when its AI-generated confidence score is higher than its real confidence score.

\mypara{Undress Detector}
To train a detector capable of identifying AI undress editing, we manually curated 103 matched pairs of original and AI-undressed images from our full dataset, yielding 206 images in total.
These images were split into a training set and a validation set at an 8:2 ratio, resulting in 164 training images and 42 validation images.
Each image was labeled as one of two classes: real (original) or fake (AI undress-edited).
We fine-tuned a pre-trained \texttt{ViT-Base-Patch16-224}~\cite{vit_base_patch16_224} for binary classification.
To mitigate overfitting on the small training set, we apply data augmentation and train with AdamW~\cite{LH19} (learning rate =$5 \times 10^{-5}$) with a cosine annealing schedule and early stopping.
Training converges in 13 epochs, achieving a validation F1 of 1.0.

\mypara{Celebrity Classifier}
We design a dual-agent celebrity classifier using \texttt{gemini-3-flash-preview}.
First, \textit{Agent~1} analyzes a cropped face and identifies the person based solely on facial features (e.g., face shape and eye configuration).
The prompt for \textit{Agent~1} is shown in~\autoref{figure:prompt_of_agent1}.
Next, \textit{Agent~2} examines the full image to verify \textit{Agent~1}'s prediction.
If \textit{Agent~1} classifies the target as a celebrity, \textit{Agent~2} uses the prompt in~\autoref{figure:prompt_of_agent2_celebrity}; otherwise, it uses the prompt in~\autoref{figure:prompt_of_agent2_non-celebrity}.

\begin{figure}[ht]
\begin{promptbox}
Please analyze this face image and tell me if there is a celebrity or famous person.

If you recognize them:
- Provide their full name
- Rate your confidence (0-100%)
- Provide a brief reason for your confidence level
- Remember only recognize as celebrity when you are very confident about the facial features such as face shape, eyes color, hair color, and other visual cues. If you are not sure, please say "Unknown".

If you don't recognize them or there's no clear face:
- Just say "Unknown"

Format your response as:

Name: [Full Name or Unknown]

Confidence: [0-100\%

Reason: for example, "The person in the image has been recognized as [Celebrity Name] with a confidence of [Confidence] based on facial features, hair color, eyes color and other visual cues."

Keep the response brief.
\end{promptbox}
\caption{Prompt of \textit{Agent~1} in celebrity classifier.}
\label{figure:prompt_of_agent1}
\end{figure}

\begin{figure}[ht]
\begin{promptbox}
Please look at this full image carefully and verify if the person in the image is really the celebrity "\{celebrity\_name\}".

Analyze both:
1. The facial features, hair, eyes to confirm it matches \{celebrity\_name\}
2. The photo context: professional photos (from TV shows, movies, or professional photoshoots) with good lighting and clean backgrounds are more likely to be celebrities, while regular people's photos may have messy backgrounds and uneven lighting

IMPORTANT: If you think the image is a deepfake or AI-generated, but the person depicted is indeed \{celebrity\_name\}, still answer "Yes". We want to classify images by who is depicted, regardless of whether the image is real or AI-generated.

Respond in this format:
Is this \{celebrity\_name\}? [Yes/No]

Confidence: [0-100]\%

Reason: [Brief explanation considering both facial match and photo context]

Be very careful and only say "Yes" if you are very confident this is indeed \{celebrity\_name\} AND the photo quality/context supports it being a celebrity photo.
\end{promptbox}
\caption{Prompt of \textit{Agent~2} if \textit{Agent~1} recognizes the target as a celebrity.}
\label{figure:prompt_of_agent2_celebrity}
\end{figure}

\subsection{Detectors and Classifiers Used in RQ3}

\mypara{Request Detector}
We design a request detector to determine whether a post is requesting SNEACI. 
First, if the post contains an image, we use the NSFW detector to classify it as NSFW or SFW. 
We then provide the post text, the first post of the thread, together with the image label when applicable, to \texttt{GPT-4o-mini}~\cite{GPT-4o-mini}, using the prompt in~\autoref{figure:prompt_of_request_detector}, to judge whether the post is a request. 
We manually verify the detector on 200 randomly sampled posts and find that it achieves 95.0\% accuracy.

\mypara{Response Matcher}
A request usually either receives a response from a fulfiller within the same thread or receives no response at all. 
Therefore, for each request, we traverse all SNEACI items posted in the same thread and determine whether any of them is a response to that request.
For requests that include an image, we judge a match based on whether the SNEACI and the request contain the same face. 
Specifically, we use InsightFace to extract facial embeddings from the request image and the candidate SNEACI, and compute the cosine similarity between them. If the similarity exceeds 0.3, we treat them as depicting the same person. 
When the candidate SNEACI is a video, we first examine the first frame for a detectable face. 
If no face is found, we randomly sample another frame, repeating this process for up to five attempts. 
If no face is detected after five tries, we treat the video as having no identifiable face.
For text-only requests, we determine whether the person described in the SNEACI matches the target described in the request text.

\mypara{Request Target Classifier}
To determine whether a request target is a celebrity or a non-celebrity, we first examine whether the request text explicitly mentions a person’s name. 
If so, we directly infer celebrity status from the name. 
Otherwise, we use the celebrity classifier to classify the target appearing in the associated request image.

\clearpage

\begin{figure}[htbp]
\begin{promptbox}
The first analysis on the cropped face image gave this result:
"\{agent1\_result\}"

Now, please look at this FULL image carefully (with background and context) and tell me if the person is a celebrity or famous person.

Analyze both:
1. The facial features, hair, eyes (consider the first analysis as reference)
2. The photo context: professional photos (from TV shows, movies, or professional photoshoots) with good lighting and clean backgrounds are more likely to be celebrities, while regular people's photos may have messy backgrounds and uneven lighting

IMPORTANT: If you think the image is a deepfake or AI-generated, but the person depicted is a celebrity, you should still recognize them. We want to classify images by who is depicted, regardless of whether the image is real or AI-generated.

Respond in this format:

Is this a celebrity? [Yes/No]

If Yes, Celebrity Name: [Full Name]

Confidence: [0-100]\%

Reason: [Brief explanation considering both facial features and photo context, and why you agree or disagree with the first analysis]

Be very careful and only say "Yes" if you are confident this is a celebrity photo.
\end{promptbox}
\caption{Prompt of \textit{Agent~2} if \textit{Agent~1} recognizes the target as a non-celebrity.}
\label{figure:prompt_of_agent2_non-celebrity}
\end{figure}

\begin{figure}[ht]
\begin{promptbox}
You are analyzing messages from a 4chan thread. Your task is to determine if a message is a REQUEST for image editing, manipulation, or generation.

Common request types include:

- Asking for nudification/clothes removal (e.g., "can you undress her", "nude edit please")

- Requesting sex/sexual edits

- Asking for specific style/pose (e.g., "in this style please", "same edit for her")

- General requests for edits (e.g., "can someone do her", "please wizard")

- Short messages with implied request (e.g., "please", "her please", "this one", "plox") when they have an image attached

- Posts with SFW (safe for work) images but minimal/no text - these are usually requests for nudification

- Request with a celebrity's name or character name

NOT requests:

- Simple acknowledgments without asking (e.g., "thanks!" as reply, "nice work", "amazing")

- Just "bump" or "bumping"

- Comments without asking for anything (e.g., "she's hot")

- Posts with NSFW images and no request text (these are likely deliveries/results, not requests)

Thread Context: \{first\_post\_of\_the\_thread\}

Message Context: \{post\}

Respond with ONLY "YES" if it's a request, or "NO" if it's not a request.
\end{promptbox}
\caption{Prompt of request detector. (To avoid further harm, we omit explicit sexual terms and celebrity names from the prompt shown in the figure; the full version is provided in the code.)}
\label{figure:prompt_of_request_detector}
\end{figure}

%-------------------------------------------------------------------------------
\end{document}